\documentstyle[prc,aps,graphicx,enumerate]{revtex}

\newcommand{\mean}[1]{\left\langle #1 \right\rangle} 
\newcommand{\connex}[1]{\left\langle #1 \right\rangle_c} 
\newcommand{\cumul}[1]{\left\langle\!\!\left\langle #1 
  \right\rangle\!\!\right\rangle}

\begin{document}

\preprint{Saclay-T00/110}
\draft

\title{A new method for measuring azimuthal distributions 
in nucleus-nucleus collisions}

\author{Nicolas Borghini,$^{1,}$
\footnote{Present address: Service de Physique Th\'eorique, CP225, 
Universit\'e Libre de Bruxelles, B-1050 Brussels}
Phuong Mai Dinh,$^2$ 
and Jean-Yves Ollitrault$^2$}
\address{$^1$~Laboratoire de Physique Th\'eorique des Particules 
El\'ementaires 
\\ 
Universit\'e Pierre et Marie Curie, 
4 place Jussieu,
F-75252 Paris cedex 05
}
\address{$^2$~Service de Physique Th\'eorique, CEA-Saclay, 
F-91191 Gif-sur-Yvette cedex}

\maketitle

\begin{abstract}
The methods currently used to measure azimuthal distributions of 
particles in heavy ion collisions assume that all azimuthal
correlations between particles result from their correlation 
with the reaction plane. However, other correlations exist, 
and it is safe to neglect them only if azimuthal anisotropies 
are much larger than $1/\sqrt{N}$, with $N$ the total number 
of particles emitted in the collision. This condition is 
not satisfied at ultrarelativistic energies. 
We propose a new method, based on a cumulant expansion of 
multiparticle azimuthal correlations, which allows measurements of 
much smaller values of azimuthal anisotropies, down to $1/N$. 
It is simple to implement and can be used to measure 
both integrated and differential flow. 
Furthermore, this method automatically eliminates the 
major systematic errors, which are due to azimuthal asymmetries 
in the detector acceptance. 
\end{abstract}


\section{Introduction}

In heavy ion collisions, much work is devoted to the study of the
azimuthal distributions of outgoing particles, and in
particular of distributions with respect to the reaction plane. 
Since these distributions reflect the interactions between 
particles, possible anisotropies, the so-called ``flow,'' reveal
information on the hot stages of the collision: thermalization,
pressure gradients, time evolution, etc.~\cite{ollitrault98}

Since the orientation of the reaction plane is not known 
{\it a priori}, flow measurements are usually extracted from 
two-particle azimuthal correlations. 
This is based on the idea that azimuthal correlations between 
two particles are generated by the correlation of the azimuth 
of each particle with the reaction plane. 
The assumption that this is the only source 
of two-particle azimuthal correlations, or at least that 
other sources can be neglected,  
dates back to the early days of the flow \cite{danielewicz85}. 
It still underlies the analyses done at 
ultrarelativistic energies, both at the Brookhaven AGS 
and the CERN SPS. 

However, we have shown in recent papers~\cite{dinh,borghini} 
that other sources of azimuthal correlations (which we refer 
to as ``nonflow'' correlations) are of comparable magnitude 
and must be taken into account in the flow analysis. 
We have studied in detail the well-known correlations 
due to global momentum conservation~\cite{danielewicz88} 
and those due to quantum correlations between identical 
particles~\cite{dinh}. We have also discussed other correlations 
due to resonance decays and final state interactions~\cite{borghini}.

Nonflow correlations scale with the total multiplicity $N$ like 
like $1/N$. Thus, they become large for peripheral collisions. 
It is important to take them into account, in particular, 
when studying the centrality dependence of the flow, which has 
been recently proposed as a sensitive probe of the phase transition 
to the quark-gluon plasma \cite{Sorge,Volo99,Bravina}. 

Clearly, a reliable flow analysis should eliminate nonflow correlations. 
Correlations due to momentum conservation can be calculated 
analytically and subtracted from the measured 
correlations, so as to isolate the correlations due to 
flow~\cite{borghini,danielewicz88}; short range 
correlations can be measured independently and 
subtracted in the same way~\cite{dinh}. 
Other well-identified nonflow correlations can be estimated 
through a Monte-Carlo simulation \cite{poskanzer98}. 
This method was used by the WA93 Collaboration to estimate 
direct correlations from $\pi^0\to\gamma\gamma$ decays~\cite{WA93}. 
Alternatively, one can attempt to eliminate nonflow correlations 
directly at the experimental level: effects of momentum conservation 
cancel if the detector used in the flow analysis is symmetric
with respect to midrapidity~\cite{danielewicz88}; short range 
correlations are eliminated if one correlates two subevents 
separated by a gap in rapidity. This is the method recently used 
by the STAR Collaboration at RHIC~\cite{STAR00}. 
In the STAR paper, the correlations between pions of the same 
charge are also compared with correlations between $\pi^+$
and $\pi^-$: correlations from $\rho^0\to\pi^+\pi^-$ are thus 
found to be negligible. 

Nevertheless, nonflow correlations remain which cannot be handled 
so simply. Correlations due to resonance decays, for instance, are 
hard to estimate (this would require a detailed knowledge 
of the collision dynamics) and 
there is no known systematic way to eliminate them at the experimental
level; more importantly, the production of minijets will contribute 
to azimuthal correlations in the experiments at higher energies, 
at RHIC and LHC. 
Finally, the existence of other sources of nonflow correlations, 
so far unknown, cannot be excluded.  

The purpose of this paper is to propose a new method for the 
flow analysis which requires no knowledge of nonflow 
correlations. The general idea is to eliminate these latter using 
higher order azimuthal correlations. 
Higher order correlations were previously used in~\cite{Jiang92} 
to show qualitatively the collectivity of flow. 
The study presented in this paper is more quantitative~: by means of a 
cumulant expansion, we are able to extract the value of the flow from 
multiparticle correlations. 
The method we propose is more reliable, and in
many respects simpler than traditional methods~\cite{poskanzer98}. 
In particular, 
detector defects, which must be considered carefully when
measuring anisotropies of a few percent, can be corrected in 
a compact and elegant way. 

In Sec.\ \ref{s:cumulants}, we give the principle 
of our method as well as orders of magnitude. We show in particular 
that this method is more sensitive: it 
allows measurements of azimuthal 
anisotropies down to values of order $1/M$, 
instead of $1/\sqrt{N}$ with the standard analysis, where $N$ denotes 
the total multiplicity of particles emitted in the collision. 

Then, we show how the method can be implemented practically. 
As usual, the measurement of azimuthal
distributions is performed in two steps. 
First, one reconstructs approximately the orientation of the reaction plane 
from the directions of many emitted particles, and one estimates 
the statistical uncertainties associated with this reconstruction. 
In fact, this first step amounts to measuring the value of the flow, 
integrated over some region of phase space (corresponding 
typically to a detector). 
We show in Sec.\ \ref{s:integrated} how this measurement can be 
done using moments of the distribution of the $Q$-vector, which 
generalizes the transverse momentum transfer introduced by 
Danielewicz and Odyniec in order to estimate 
the azimuth of the reaction plane~\cite{danielewicz85}. 
We also discuss an improved version of 
the subevent method introduced by the same authors to 
estimate the accuracy of the reaction plane reconstruction. 

The second step in the flow analysis is to perform more detailed 
measurements of azimuthal distributions, 
for various particles, as a function of rapidity and/or transverse 
momentum. We refer to these detailed measurements as to 
``differential flow''. 
They are usually performed by measuring distributions 
with respect to the reconstructed reaction plane, and then correcting
for the statistical errors in this reconstruction, which have been 
estimated previously. 
Here, the differential flow will be extracted directly from the 
correlation between the azimuths of the outgoing particles and the 
$Q$-vector, as explained in Sec.\ \ref{s:differential}.
The discussion applies so far to an ideal detector. 
A general way of implementing acceptance corrections 
adapted to our method is discussed in Sec.\ \ref{s:acceptance}. 
Finally, the correct procedure is summarized in Sec.\ 
\ref{s:summary}. 
Readers already familiar with flow analysis and willing 
to apply our method may go directly to this last section.

\section{Cumulant expansion of azimuthal correlations}
\label{s:cumulants}

As the standard methods of flow analysis~\cite{poskanzer98}, 
our method is based on a Fourier expansion of azimuthal distributions
\cite{voloshin96} which is defined in Sec.\ \ref{s:fourier}. 
Then, in Sec.\ \ref{s:twoparticle}, we discuss two-particle 
azimuthal correlations, on which the standard flow analysis 
relies, and show that they decompose into a contribution from 
flow and an additional term of order $1/N$ which corresponds 
to nonflow correlations; this latter contribution limits the 
sensitivity of the traditional method.  
In Sec.\ \ref{s:multiparticle}, the decomposition is 
generalized to multiparticle correlations. 
Finally, in Sec.\ \ref{s:multiparticle2}, we show that 
this decomposition of multiparticle correlations allows 
us to obtain more sensitive measurements of flow.

\subsection{Fourier coefficients}
\label{s:fourier}

We call ``flow'' the azimuthal correlations between 
the outgoing particles and the reaction plane. 
These are conveniently characterized in terms of the Fourier 
coefficients $v_n$~\cite{voloshin96} which we now define. 
In most of this paper, we shall work with 
a coordinate system in which the $x$ axis is the impact 
direction, and $(x,z)$ the reaction plane, while  
$\phi$ denotes the azimuthal angle with respect to the
reaction plane. In this frame, the momentum of a particle of mass
$m$ is 
\begin{equation}
\label{eq1}
{\bf p}=\pmatrix{p_x= p_T\cos\phi\hfill\cr p_y=p_T\sin\phi\hfill\cr
p_z=\sqrt{p_T^2+m^2}\sinh y\hfill},
\end{equation}
where $p_T$ is the transverse momentum and $y$ the rapidity. 
Since the orientation of the reaction plane is unknown in 
experiments, so is the azimuth $\phi$. Therefore $p_x$ and $p_y$ 
are not measured directly. 

When necessary, we shall 
denote by $\bar\phi$ the azimuthal angle in the 
laboratory frame.  Unlike $\phi$, $\bar\phi$ is a measurable quantity, 
related to $\phi$ by 
$\bar\phi=\phi+\phi_R$, where $\phi_R$ is the unknown azimuthal angle 
of the reaction plane in the laboratory system. 

With these definitions, $v_n$ can be expressed as a
function of the one-particle momentum distribution 
$f({\bf p})\equiv {\rm d}N/{\rm d}^3{\bf p}$ 
\begin{equation}
\label{vn}
v_n({\cal D})\equiv\mean{e^{i n\phi}} =
{\displaystyle\int_{\cal D} e^{i n\phi} f({\bf p}) \, {\rm d}^3{\bf p}\over
\displaystyle \int_{\cal D}f({\bf p})\,{\rm d}^3{\bf p}},
\end{equation}
where the brackets denote an average value over many events, 
and ${\cal D}$ represents a phase space window in the $(p_T,y)$ plane 
where flow is measured, typically corresponding to a detector. 
Since the particle source is symmetric with respect to the reaction plane
for spherical nuclei, $\mean{\sin n\phi}$ vanishes and $v_n$ is real. 

The purpose of the flow analysis is to extract $v_n$ from the data. 
Only the first two coefficients $v_1$ and $v_2$ have been published. 
They are usually called directed and elliptic flow, 
respectively. 
There are so far very few measurements of higher order coefficients. 
The E877 experiment at the Brookhaven AGS reported values compatible 
with zero for $v_3$ and $v_4$ \cite{barrette94}. 
Nonvanishing values of higher harmonics, up to $v_6$, were reported 
from preliminary analyses 
at the CERN SPS \cite{lenkeit,NA49unpub}. However, the latter results 
are likely to be strongly biased by quantum two-particle correlations
\cite{dinh}. 
At the energies of the CERN SPS, $v_1$ and $v_2$ are of the order
of a few percent \cite{na49}, close to the limit of detectability 
with the standard methods, hence the need for a new, more sensitive 
method.

\subsection{Two-particle correlations}
\label{s:twoparticle}

Since the actual orientation of the reaction plane is not known 
experimentally, one can 
only measure relative azimuthal angles between outgoing 
particles. 
The standard flow analysis relies on the measurement of two-particle 
azimuthal correlations, which involve the two-particle distribution 
$f({\bf p}_1,{\bf p}_2)={\rm d}N/{\rm d}^3{\bf p}_1{\rm d}^3{\bf p}_2$:
\begin{equation}
\label{defcn}
\mean{e^{in(\phi_1-\phi_2)}}_{{\cal D}_1\times{\cal D}_2} = 
{\displaystyle\int_{{\cal D}_1\times{\cal D}_2}
\displaystyle{e^{in(\phi_1-\phi_2)}
{\displaystyle f({\bf p}_1,{\bf p}_2) \,
{\rm d}^3{\bf p}_1{\rm d}^3{\bf p}_2}}\over
{\displaystyle\int_{{\cal D}_1\times{\cal D}_2}
{\displaystyle f({\bf p}_1,{\bf p}_2) \,
{\rm d}^3{\bf p}_1{\rm d}^3{\bf p}_2}}}.
\end{equation}
The standard analysis neglects nonflow correlations. 
Under that assumption, the two-particle momentum distribution factorizes:
\begin{equation}
f({\bf p}_1,{\bf p}_2) = f({\bf p}_1)f({\bf p}_2).
\end{equation}
Then, Eqs.\ (\ref{vn}) and (\ref{defcn}) give 
\begin{equation}
\label{flo1}
\mean{e^{in(\phi_1-\phi_2)}}_{{\cal D}_1\times{\cal D}_2} =
v_n({\cal D}_1)v_n({\cal D}_2).
\end{equation}
This equation means that the only azimuthal correlation between 
two particles results from their correlation with the 
reaction plane. 
Measuring the left-hand side of Eq.\ (\ref{flo1}) in various phase 
space windows, one can then reconstruct $v_n$ from this equation, 
up to a global sign. 
For instance, the E877 Collaboration uses 
the correlations between three rapidity windows to extract flow from 
their data~\cite{E877}.

However, nonflow correlations do exist. The two-particle distribution 
can generally be written as 
\begin{equation}
\label{fc2}
f({\bf p}_1,{\bf p}_2)=f({\bf p}_1)f({\bf p}_2)+f_c({\bf p}_1,{\bf p}_2),
\end{equation}
where $f_c({\bf p_1},{\bf p_2})$ denotes the correlated part of the 
distribution. 
There are various sources of such correlations, among which 
global momentum conservation, resonance decays (in which 
the decay products are correlated), final state Coulomb, strong 
or quantum interactions~\cite{dinh,borghini}. 

In the coordinate system we have chosen, where the reaction 
plane is fixed, $f_c({\bf p_1},{\bf p_2})$ is typically of 
order $1/N$ relative to the uncorrelated part, where $N$ is 
the total number of particles emitted in the collision. 
This order of magnitude can easily be understood in the case 
of correlations between decay products, such as $\rho\to\pi\pi$. 
A significant fraction of the pions produced in a heavy ion collision 
originate from this decay, and the conservation of energy and momentum 
in the decay gives rise to a large correlation between the reaction 
products. Since a large number of $\rho$ mesons are produced 
in a high energy nucleus-nucleus collision, 
the probability that two arbitrary pions originate from the same $\rho$
is of order $1/N$. 
This $1/N$ scaling also holds for the correlation due to global 
momentum conservation~\cite{borghini,danielewicz88}. 

Inserting Eq.\ (\ref{fc2}) in expression (\ref{defcn}), one finds, instead
of Eq.\ (\ref{flo1}),  
\begin{equation}
\label{c2}
\mean{e^{i n(\phi_1-\phi_2)}}_{{\cal D}_1\times{\cal D}_2} =
v_n({\cal D}_1)v_n({\cal D}_2)
+ \connex{e^{i n(\phi_1-\phi_2)}}.
\end{equation}
The left-hand side represents the measured two-particle 
azimuthal correlation. The first term in the right-hand 
side is the contribution of flow to this correlation,
while the second term $\connex{e^{i n(\phi_1-\phi_2)}}$ 
denotes the contribution of the correlated part $f_c$.
The latter term corresponds to azimuthal correlations 
which do not arise from flow: we call them ``direct'' correlations, 
in opposition to the indirect correlations arising from the 
correlation with the reaction plane, that is, from flow. 

Since the correlated two-particle distribution 
$f_c({\bf p}_1,{\bf p}_2)$ is of order $1/N$, so is the 
second term in the right-hand side of Eq.\ (\ref{c2}), 
which therefore reads 
\begin{equation}
\label{c2order}
\mean{e^{i n(\phi_1-\phi_2)}}_{{\cal D}_1\times{\cal D}_2} =
v_n({\cal D}_1)v_n({\cal D}_2)
 + O\left({1\over N}\right).
\end{equation}
However, one must be careful with this order of magnitude. 
Strictly speaking, it holds only when momenta are averaged 
over a large region of phase space. In the case of the 
short range correlations due to final state interactions
(Coulomb, strong, quantum) the correlations vanish as soon as 
the phase spaces ${\cal D}_1$ and ${\cal D}_2$ of the two particles 
are widely separated. This is the method 
used in~\cite{STAR00} to get rid of such correlations. 
If, on the other hand, ${\cal D}_1$ and ${\cal D}_2$ coincide, 
the short range correlations are larger than expected from 
Eq.\ (\ref{c2order}): in this equation, the total number 
of emitted particles $N$ should be replaced by the number 
of particles $M$ used in the flow analysis, which is 
smaller in practice. 
Furthermore, in the case of correlations due to the quantum 
(HBT) effect, the nonflow correlation scales like $1/N$ 
only if the source radius $R$ scales like 
$N^{1/3}$~\cite{dinh,borghini}.   
From now on, we shall omit the subscript ${\cal D}$ for sake of 
brevity. Note, however, that all the averages we shall consider 
are over a region of phase space which is not necessarily the 
whole space, but may be restricted to the $(p_T,y)$ acceptance 
of a detector. 
This will be especially important in Sec.\ \ref{s:acceptance}, when 
we discuss acceptance corrections. 

Equation (\ref{c2order}) shows that  nonflow correlations can be 
neglected if $v_n\gg N^{-1/2}$.
At SPS energies, the flow is weak and this condition is not fulfilled.
Indeed, we have shown~\cite{dinh,borghini} 
that the values of flow measured by the NA49 Collaboration 
at CERN are considerably modified once nonflow correlations are taken 
into account.

\subsection{Multiparticle correlations and the cumulant expansion}
\label{s:multiparticle}

The failure of the standard analysis is due to the impossibility to 
separate the correlated part from the 
uncorrelated part in Eq.\ (\ref{fc2}) at the level of two-particle 
correlations. The main idea of this paper is to perform this separation 
using multiparticle correlations. 
The decomposition of the particle distribution into correlated and 
uncorrelated parts in Eq.\ (\ref{fc2}) can be generalized to an 
arbitrary number 
of particles. For instance, the three-particle distribution can be 
decomposed as 
\begin{eqnarray}
\label{fc3}
\frac{{\rm d}N}{{\rm d}{\bf p}_1{\rm d}{\bf p}_2{\rm d}{\bf p}_3} \equiv
f({\bf p}_1,{\bf p}_2,{\bf p}_3)&=&f_c({\bf p}_1)f_c({\bf p}_2)f_c({\bf p}_3)
\cr
&+&f_c({\bf p}_1,{\bf p}_2)f_c({\bf p}_3)
+f_c({\bf p}_1,{\bf p}_3)f_c({\bf p}_2) 
+f_c({\bf p}_2,{\bf p}_3)f_c({\bf p}_1)\cr
&+&f_c({\bf p}_1,{\bf p}_2,{\bf p}_3),
\end{eqnarray}
where $f_c({\bf p}_1)\equiv f({\bf p}_1)$. The last term 
$f_c({\bf p}_1,{\bf p}_2,{\bf p}_3)$ corresponds to the genuine 
three-particle correlation, which is of order $1/N^2$. 

To understand this  order of magnitude, let us take a simple 
example: the $\omega$ meson decays mostly into three pions. 
First of all, this decay generates direct two-particle correlations: 
the relative momentum between any two of the outgoing 
pions is restricted by energy and momentum conservation. 
The corresponding correlation is of order $1/N$ as 
discussed previously in the case of $\rho\to\pi\pi$ decays. 
It corresponds to the second, third and fourth term in the 
right-hand side of Eq.\ (\ref{fc3}). 
As stated above, the last term in this equation stands for the direct 
three-particle correlation. 
The corresponding correlation between the decay products 
of a given $\omega$ is 
of order unity, while the probability that three arbitrary pions 
come from the same $\omega$ scales with $N$ like $1/N^2$. 
Thus the correlation between three random pions is of order $1/N^2$.

More generally, the decomposition of the $k$-particle distribution 
yields a correlated part $f_c({\bf p}_1,\dots,{\bf p}_k)$ of order 
$1/N^{k-1}$. 
Generalizing the above discussion of $\omega\to\pi\pi\pi$ decay, 
the decay of a cluster of $k$ particles will generate correlations 
$f_c({\bf p}_1,\dots,{\bf p}_{k'})$ with $k'\le k$. 
For instance, momentum conservation, which is an effect involving all 
$N$ particles emitted in a collision, produces direct $k$-particle 
correlations for arbitrary $k$. 

Such a decomposition is similar to the cluster expansion which is 
well known in the theory of imperfect gases~\cite{hill}. 
In the language of probability theory, this is known as the 
cumulant expansion~\cite{vanKampen}. 
Equations (\ref{fc2}) and (\ref{fc3}) can be represented diagrammatically 
by Figs. \ref{fig:fig2p} and \ref{fig:fig3p}. 
In these figures, correlated distributions $f_c$
are represented by enclosed sets of points,  
i.e.\ they correspond to connected diagrams. 
\begin{center}
\begin{figure}[ht!]
\centerline{\includegraphics*[width=0.28\linewidth]{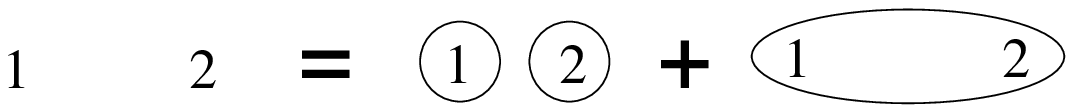}}
\medskip
\caption{Decomposition of the two-particle distribution into 
uncorrelated and correlated components. The second term in the 
right-hand side is smaller than the first by a factor of order 
$1/N$.}
\label{fig:fig2p}
\end{figure}
\begin{figure}[ht!]
\centerline{\includegraphics*[width=0.65\linewidth]{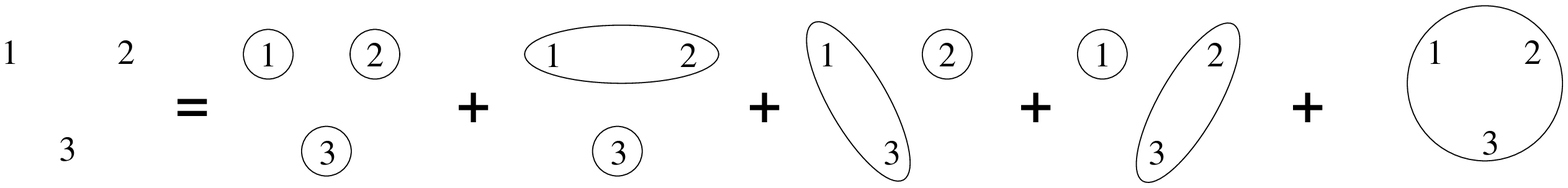}}
\medskip
\caption{Decomposition of the three-particle distribution.
The last term in the right-hand side is of order $1/N^2$ relative 
to the first, while the three remaining terms are of relative order $1/N$.}
\label{fig:fig3p}
\end{figure}
\end{center}
More generally, in order to decompose the $k$-point function 
$f({\bf p}_1,\cdots,{\bf p}_k)$, 
one first takes all possible partitions of the set of points 
$\{{\bf p}_1,\cdots,{\bf p}_k\}$. 
To each subset of points $\{{{\bf p}_i}_1,\cdots,{{\bf p}_i}_m\}$, 
one associates the corresponding correlated function 
$f_c({{\bf p}_i}_1,\cdots,{{\bf p}_i}_m)$. 
The contribution of a given partition is
the product of the contributions of each subset. 
Finally, $f({\bf p}_1,\cdots,{\bf p}_k)$ is the 
sum of the contributions of all partitions. 

The equations expressing the $k$-point functions $f$ in terms of 
the correlated functions $f_c$ can be inverted order by order, so as to 
isolate the term of smallest magnitude: 
\begin{eqnarray}
f_c({\bf p}_1)&=&f({\bf p}_1)\cr
f_c({\bf p}_1,{\bf p}_2)&=&f({\bf p}_1,{\bf p}_2)-f({\bf p}_1)f({\bf p}_2)\cr
f_c({\bf p}_1,{\bf p}_2,{\bf p}_3)&=&
f({\bf p}_1,{\bf p}_2,{\bf p}_3)
-f({\bf p}_1,{\bf p}_2)f({\bf p}_3)-f({\bf p}_1,{\bf p}_3)f({\bf p}_2)-f({\bf p}_2,{\bf p}_3)f({\bf p}_1)
+2f({\bf p}_1)f({\bf p}_2)f({\bf p}_3).
\end{eqnarray}

The cumulant expansion has been used previously in high energy 
physics to characterize multiparticle correlations: 
it has been applied to correlations in rapidity~\cite{carruthers89}
and to Bose-Einstein quantum correlations~\cite{giovannini77,eggers93}. 
In these studies, the interest was mainly in short range correlations. 
The use of higher order cumulants was therefore limited by 
statistics: the probability that three or more particles are very 
close in phase space is small. 
In this paper, we are interested in collective flow, which by 
definition produces a long range correlation, so that the limitation 
due to statistics is not so drastic. It will indeed be shown in Sec.\ 
\ref{s:statistical} that cumulants up to order 6 can be measured, 
depending on the event multiplicity and available statistics. 

We shall deal with multiparticle azimuthal 
correlations, which generalize the two-particle azimuthal
correlations in Eq.\ (\ref{c2}), and can be decomposed
in the same way. Referring to the diagrammatic representation in 
Figs. \ref{fig:fig2p} and \ref{fig:fig3p}, we shall name the
contribution of $f_c({\bf p}_1,\cdots,{\bf p}_k)$ to an azimuthal 
correlation, i.e.\ the genuine $k$-particle correlation, the ``connected 
part'' of the correlation or, equivalently, the ``direct'' $k$-particle 
correlation. 

\subsection{Measuring flow with multiparticle azimuthal correlations}
\label{s:multiparticle2}

Our method, which we now explain, allows the detection of small 
deviations from an isotropic distribution. 
If the source is isotropic, there is no flow, 
and the orientation of the reaction plane 
does not influence the particle distribution. We can therefore 
consider that the reaction plane has a fixed direction 
in the laboratory coordinate system, so that
the cumulant expansion can be performed in that frame: in 
other terms, we replace $\phi$ by the measured azimuthal angle $\bar\phi$. 
One then measures the $k^{\rm th}$ cumulant of the multiparticle 
azimuthal correlation, which is of order $N^{1-k}$ if the 
distribution is isotropic. Flow will appear as a deviation 
from this expected behaviour. 

Let us be more explicit. We are dealing with azimuthal correlations. 
When the source is isotropic, that is, if the $k$-particle distribution 
remains unchanged when all azimuthal angles are shifted by the same quantity 
$\alpha$, the flow coefficients (\ref{vn}) obviously vanish. 
Therefore, the two-particle azimuthal correlation (\ref{c2})
reduces to its connected part, of order $1/N$. 
As a further consequence of isotropy, averages like 
$\mean{e^{in(\phi_1+\phi_2-\phi_3)}}$ vanish: only 
$2k$-particle azimuthal correlations involving $k$ powers of $e^{in\phi}$ 
and $k$ powers of $e^{-in\phi}$ are nonvanishing. 
For instance, the four-particle correlation 
$\mean{e^{in(\phi_1+\phi_2-\phi_3-\phi_4)}}$ is {\it a priori} non
vanishing. Introducing the cumulant expansion defined in 
Sec.\ \ref{s:multiparticle}, this correlation can be decomposed into 
\begin{eqnarray}
\label{c4}
\mean{e^{in(\phi_1+\phi_2-\phi_3-\phi_4)}} & = &
\connex{e^{in(\phi_1-\phi_3)}} 
\connex{e^{in(\phi_2-\phi_4)}} +
\connex{e^{in(\phi_1-\phi_4)}} 
\connex{e^{in(\phi_2-\phi_3)}} +
\connex{e^{in(\phi_1+\phi_2-\phi_3-\phi_4)}} \cr
& = & \mean{e^{in(\phi_1-\phi_3)}} 
\mean{e^{in(\phi_2-\phi_4)}} +
\mean{e^{in(\phi_1-\phi_4)}} 
\mean{e^{in(\phi_2-\phi_3)}} +
\connex{e^{in(\phi_1+\phi_2-\phi_3-\phi_4)}}. 
\end{eqnarray}
Note that most terms in the cumulant expansion disappear 
as a consequence of isotropy. 
The first two terms in the right-hand side of Eq.\ (\ref{c4}) 
are products of direct two-particle correlations, and are therefore of 
order $1/N^2$, while the last term, which corresponds to the direct 
four-particle correlation, is much smaller, of order $1/N^3$. 
However, in the case of short range correlations,  it may rather be of 
order $1/M^3$, where $M$ is the number of particles used in the flow 
analysis, for the same reasons as discussed in 
Sec.\ \ref{s:twoparticle}. 

We name this latter term the ``cumulant'' to order 4 and 
denote it by $\cumul{e^{in(\phi_1+\phi_2-\phi_3-\phi_4)}}$. 
Using Eq.\ (\ref{c4}), it can be expressed as a function 
of the measured two- and four-particle azimuthal correlations: 
\begin{equation}
\label{defcumul4}
\cumul{e^{in(\phi_1+\phi_2-\phi_3-\phi_4)}} \equiv 
\mean{e^{in(\phi_1+\phi_2-\phi_3-\phi_4)}} -
\mean{e^{in(\phi_1-\phi_3)}} \mean{e^{in(\phi_2-\phi_4)}} -
\mean{e^{in(\phi_1-\phi_4)}} \mean{e^{in(\phi_2-\phi_3)}}. 
\end{equation}
The reason why we introduce a new notation here is that the 
cumulant to order 4 will always be defined by (\ref{defcumul4}) 
in this paper, even when the source is not isotropic. 
Now, if the source is not isotropic, the decomposition 
of the four-particle azimuthal correlation involves many terms 
which have been omitted in Eq.\ (\ref{c4}) (see Appendix 
\ref{s:app-c4c}), so that the cumulant 
$\cumul{e^{in(\phi_1+\phi_2-\phi_3-\phi_4)}}$ 
no longer corresponds to the connected 
part $\connex{e^{in(\phi_1+\phi_2-\phi_3-\phi_4)}}$. 

In the isotropic case, the cumulant 
$\cumul{e^{in(\phi_1+\phi_2-\phi_3-\phi_4)}}$ 
involves only direct four-particle correlations: 
the two-particle correlations have been eliminated 
in the subtraction. 
In order to illustrate this statement, let us consider 
two decays $\rho \to \pi\pi$, and ``turn off'' all 
other sources of azimuthal correlations. 
We label 1 and 2 the pions emitted by the 
first resonance, 3 and 4 the pions emitted by the second. 
There are correlations between $\pi_1$ and $\pi_2$, 
or between $\pi_3$ and $\pi_4$, so that the measured 
four-particle correlation, i.e.\ the left-hand side of 
Eq.\ (\ref{c4}), does not vanish. 
However, there is no {\em direct} four-particle correlation between the 
four outgoing pions, so that the cumulant (\ref{defcumul4}) vanishes. 
More generally, if particles are produced in clusters of $k$
particles, there are measured azimuthal correlations to all orders, 
but the cumulants to order $k'>k$ vanish. 

Let us now consider small deviations from isotropy, i.e.\ weak flow. 
The two-particle azimuthal correlation receives 
a contribution $v_n^2$ according to Eq.\ (\ref{c2}).
For similar reasons, the four-particle correlation 
gets a contribution $v_n^4$. 
The cumulant defined by Eq.\ (\ref{defcumul4}) thus becomes 
(see Appendix \ref{s:app-c4c})
\begin{equation}
\label{c4order}
\cumul{e^{in(\phi_1+\phi_2-\phi_3-\phi_4)}} =
  -v_n^4 + O\left({1\over N^3} + \frac{v_{2n}^2}{N^2}\right)
\end{equation}
where the coefficient $-1$ in front of $v_n^4$ is found by 
replacing each factor $e^{in\phi}$ or $e^{-in\phi}$ in the left-hand side 
with its average value $v_n$. The flow $v_n$ can thus 
be obtained, up to a sign, from the measured two- and four-particle
azimuthal correlations, with a better accuracy than when using only 
two-particle correlations, as we shall see shortly. 

It should be noticed that the cumulant involves a contribution from the 
higher order harmonic $2n$, of magnitude $v_{2n}^2/N^2$. 
This contribution does not interfere with the measurement of $v_n$
provided the following condition is satisfied:
\begin{equation}
\label{limitv2}
|v_{2n}|\ll Nv_n^2.
\end{equation}
Since $v_n$ is measurable only if $v_n\gg 1/N$, as we shall 
see later in this section, the interference with the 
harmonic $2n$ occurs only if $|v_{2n}|\gg |v_n|$. 
In practice, the only situation where this might be a problem  
is when measuring the directed flow $v_1$ at ultrarelativistic energies,
where elliptic flow $v_2$ is expected to be larger than $v_1$. 
On the other hand, this interference will not endanger the 
measurement of $v_2$, since $v_4$ should be much smaller. 

In the following, we shall always assume that condition (\ref{limitv2}) 
is fulfilled. 
Then, using Eq.\ (\ref{c4order}), it becomes possible to measure the 
flow $v_n$ as soon as it is much larger than $N^{-3/4}$. 
The sensitivity is better than with the traditional methods using 
two-particle correlations which, as we have seen, require $v_n\gg N^{-1/2}$. 

Similarly, using $2k$-particle azimuthal correlations and taking the 
cumulant, i.e.\ isolating the 
connected part (which amounts to getting rid of nonflow correlations 
of orders less than $2k$), one obtains a quantity which is of magnitude 
$N^{1-2k}$ for an isotropic source. Flow gives a contribution 
of magnitude $v_n^{2k}$. 
The contribution of higher order harmonics $v_{kn}$ can be neglected
as soon as 
\begin{equation}
\label{limitvk}
|v_{kn}|\ll N^{k-1}v_n^k. 
\end{equation}
If $|v_n|\gg 1/N$, this is not a problem, unless $|v_{kn}|\gg |v_n|$. 
This is unlikely to occur, since 
one expects $v_n$ to decrease rapidly with $n$. 
Neglecting higher order harmonics, there remains the contributions 
of flow, of magnitude $v_n^{2k}$, and of direct $2k$-particle correlations, 
of magnitude $N^{1-2k}$. 
Therefore, $2k$-particle azimuthal correlations 
allow measurements of $v_n$ if it is larger than 
$N^{-1+1/2k}$. 
Since $k$ is arbitrarily large, one can ideally measure $v_n$ down to 
values of order $1/N$, instead of $1/\sqrt{N}$ with the standard methods. 
A necessary condition for the flow analysis is therefore
\begin{equation}
\label{limitv}
v_n\gg {1\over N}
\end{equation}
which will be assumed throughout this paper. 
As we shall see in Sec.\ \ref{s:statistical}, the sensitivity 
is in fact limited experimentally by statistical errors due to the finite 
number of events. 

In practice, the cumulants of multiparticle azimuthal correlations 
will be extracted from moments of the distribution of the $Q_n$-vector 
introduced in next section.

\section{Integrated flow}
\label{s:integrated}

In this section, we show how it is possible to measure the value of $v_n$ 
integrated over a phase space region. 
This measurement will serve as a reference when we perform more 
detailed measurements of azimuthal anisotropies, in Sec.\ \ref{s:differential}. 
We first define in Sec.\ \ref{s:qvector} a simple 
version of the $Q_n$-vector, 
or event flow vector, which is used in the standard flow analysis 
to estimate the orientation of the reaction plane. 
We then show, in Sec.\ \ref{s:integrated2},
that the integrated value of the flow can 
be obtained from the moments of the $Q_n$ distribution: 
eliminating nonflow correlations up to order $2k$ by 
means of a cumulant expansion, we obtain an accuracy on 
the integrated $v_n$ of magnitude $N^{-1+1/2k}$, better than 
the accuracy of standard methods if $k>1$. 
Instead of using a single event vector $Q_n$, one can do 
a similar analysis using subevents (Sec.\ \ref{s:subevents}). 
Since the order $2k$ of the calculation is 
arbitrary, we obtain with either method 
an infinite set of equations to determine $v_n$. 
The order $2k$ which should be chosen when analyzing experimental data
depends on the number of events available (Sec.\ \ref{s:statistical}). 
More general forms of the $Q_n$-vector, which allow an optimal 
flow analysis, are discussed in Sec.\ \ref{s:weighted}.
Finally, in Sec.\ \ref{s:gaussian}, we recover, as a limiting case, 
the results obtained in the limit of large multiplicity where the 
distribution of $Q_n$ is Gaussian \cite{voloshin96,ollitrault92}. 
In the whole section, we assume the analysis is performed using a 
perfectly isotropic detector; corrections to this ideal case will be 
dealt with in Sec.\ \ref{s:acceptance}.

\subsection{The $Q$-vector}
\label{s:qvector}

\subsubsection{Definition}
\label{s:sub3a1}

Consider a collision in which $M$ particles are detected with 
azimuthal angles $\phi_1,\cdots,\phi_M$. 
In order to detect possible anisotropies of the $\phi$ 
distribution, it is natural to construct an observable 
which involves all the $\phi_j$, i.e.\ a global 
quantity. For the study of the $n^{\rm th}$ harmonic, 
one uses the $n^{\rm th}$ transverse event flow 
vector \cite{poskanzer98}, 
which we write as a complex number
\begin{equation}
Q_n = \frac{1}{\sqrt{M}} \sum_{j=1}^M e^{in\phi_j},
\label{qn}
\end{equation}
where $\phi_j$ denotes the azimuthal angle of the $j^{\rm th}$ 
particle with respect to the reaction plane. 

For simplicity, we have associated a unit weight with 
each particle in Eq.\ (\ref{qn}). 
The generalization of our results to arbitrary weights 
is straightforward and will be given in Sec.\ \ref{s:weighted}. 
The $Q_n$-vector generalizes to arbitrary harmonics 
the transverse 
momentum transfer introduced by Danielewicz and 
Odyniec~\cite{danielewicz85}, which corresponds to $n=1$
and the transverse sphericity tensor introduced 
in \cite{ollitrault92,wilson92}, which corresponds to the case $n=2$. 

In practice, the number of particles  $M$ 
used for the flow analysis is 
not equal to the total multiplicity $N$ of particles 
produced in the collision, since all particles are not detected. 
However, $M$ should be taken as large as possible. 
In this paper, we shall assume that $M$ and $N$ are of the 
same order of magnitude. 
The factor $1/\sqrt{M}$ in front of Eq.\ (\ref{qn}), 
which does not appear in previous definitions of the flow 
vector \cite{danielewicz85,voloshin96}, will be explained 
in Sec.\ \ref{s:sub3a3}.

\subsubsection{Flow versus nonflow contributions}
\label{s:sub3a2}

A nonvanishing value for the average value of the flow vector, 
$\mean{Q_n }$, signals 
collective flow. Indeed, using Eqs.\ (\ref{vn}) and (\ref{qn}), 
it is related to the Fourier coefficient $v_n = \mean{e^{in\phi} }$ by
\begin{equation}
\mean{Q_n } = \sqrt{M} v_n.
\label{v-Q}
\end{equation}
Note that $\mean{Q_n}$ is real, as is $v_n$, due to the 
symmetry with respect to the reaction plane. 

As stated before, the purpose of the flow analysis is to measure $v_n$, i.e.\ 
$\mean{Q_n}$. 
This is not a trivial task because the 
azimuth of the reaction plane is unknown, so that 
the phase of $Q_n$ is unknown. 
The only measurable quantity is $|Q_n|$, the length of $Q_n$. Its 
square $Q_n Q_n^*$, where $Q_n^*$ denotes the 
complex conjugate, only depends on relative 
azimuthal angles:
\begin{equation}
\label{qq*}
Q_n Q_n^*={1\over M}\sum_{j,k=1}^M e^{in(\phi_j-\phi_k)}.
\end{equation}
In Sec.\ \ref{s:integrated2}, we shall see that the flow 
can be deduced from the moments of the distribution of 
$|Q_n|^2$, i.e.\ from the average values $\mean{|Q_n|^{2k}}$,
where $k$ is a positive integer. 
To illustrate how flow enters these expressions, 
we discuss here the second order moment 
$\mean{|Q_n|^2}$. 
Averaging Eq.\ (\ref{qq*}) over many events and using Eq.\ (\ref{c2}), 
one obtains 
\begin{equation}
\label{q2}
\mean{|Q_n|^2} = 
{1\over M}\left[M+M(M-1)
\left(v_n^2+\connex{e^{in(\phi_j-\phi_k)}}\right)\right]. 
\end{equation}
The first term corresponds to the diagonal terms $j=k$, 
i.e.\ to ``autocorrelations''. 
If there are no azimuthal correlations (neither flow nor nonflow), 
only this term remains and the average value of $|Q_n|^2$ is exactly
1. 
The second term corresponds to $j\not= k$, i.e.\ to the two-particle 
azimuthal correlations discussed in Sec.\ \ref{s:twoparticle}. 
Since $\connex{e^{in(\phi_1-\phi_2)}}$ is at most of order $1/M$, 
direct correlations give a contribution which is {\it a priori\/} 
of the same order of magnitude as autocorrelations, although 
it may be smaller in practice. 
Equation (\ref{q2}) can thus be written 
\begin{eqnarray}
\mean{|Q_n|^2} &=& 
M\left[ v_n^2+ {1 \over M} + O\left({1\over M}\right)\right]\cr
&=&\mean{Q_n}^2+1+O(1).
\label{q2order}
\end{eqnarray}
As expected from the discussion of Sec.\ \ref{s:twoparticle}, 
since $\mean{|Q_n|^2}$ involves two-particle correlations, 
flow measurements based on $\mean{|Q_n|^2}$ are reliable 
only if $|v_n|\gg 1/\sqrt{M}$. 
Smaller values of flow can be obtained 
using higher moments of the distribution of $|Q_n|^2$, 
as explained in  Sec.\ \ref{s:integrated2}. 

If flow is strong enough, the event flow vector can be used to 
estimate the orientation of the reaction plane. 
Indeed, if $|v_n|\gg 1/\sqrt{M}$, Eqs.\ (\ref{v-Q}) and
(\ref{q2order}) show that 
$Q_n\simeq \mean{Q_n}=\sqrt{M}v_n$. 
Then the phase of $Q_n$ is approximately $0$ if $v_n>0$
and $\pi$ if $v_n<0$. 
Experimentally, one defines $Q_n$ as in (\ref{qn}), with 
the azimuthal angles $\phi_j$ measured with respect to a fixed
direction in the laboratory (rather than the reaction plane, 
which is unknown). Then the azimuthal angle of the reaction 
plane $\phi_R$ can be estimated from the phase of $Q_n$, 
which we write $n\phi_Q$: $\phi_R\simeq\phi_Q$ 
(resp.\ $\phi_R\simeq\phi_Q+\pi/n$)  modulo $2\pi/n$ if 
$v_n>0$ (resp.\ $v_n<0$).

\subsubsection{Varying the centrality}
\label{s:sub3a3}

Let us now explain the factor $1/\sqrt{M}$ in the definition (\ref{qn}). 
This factor was introduced independently by 
A.~Poskanzer and S.~Voloshin, and in \cite{ollitrault95}.
It is important when 
using events with different multiplicities $M$ in the flow 
analysis, i.e. events with different centralities. 
This is the case in practice: one takes all events in 
a given centrality interval in order to increase the 
available statistics. 

If there is no flow, Eq.\ (\ref{q2order}) shows that 
$\mean{|Q_n|^2}$ is independent of $M$ since nonflow 
correlations scale like $1/M$. 
This can be understood simply: 
the sum in Eq.\ (\ref{qn}) is a random walk of $M$ unit steps,
therefore it has a length of order $\sqrt{M}$, which cancels out 
with the factor $1/\sqrt{M}$ in front. 
Flow, on the other hand, depends strongly on centrality 
(it vanishes for central and very peripheral collisions): 
according to Eq.\ (\ref{q2order}), 
it gives a positive contribution to $\mean{|Q_n|^2}$ 
which strongly depends on $M$. This allows to disentangle 
flow and nonflow effects. 

Note that flow can be detected by studying the variation of 
$\mean{|Q_n|^2}$ with centrality. This is the method 
used in~\cite{na50}: one expects $\mean{|Q_n|^2}$ 
to be minimum for the most peripheral collisions where 
the density of particles is too small for collective behaviour
to set in, and for central collisions where $v_n$ also 
vanishes from azimuthal symmetry. 
However, such a method does not allow an accurate measurement 
of flow: it is impossible to select true (i.e.\ with $b=0$) 
central collisions experimentally, 
and there may still be some flow up to large impact parameters, 
as suggested by hydrodynamic calculations in the case of 
elliptic flow \cite{ollitrault92}, and by recent  measurements 
\cite{poskanzer99}.

The method presented in this paper is more powerful in the sense 
that it allows flow measurements for a given centrality. 
The error on the centrality selection (due to the fact that 
one always selects events within a finite range of impact 
parameters) is compensated by the factor $1/\sqrt{M}$ in the 
definition of $Q_n$. 

\subsection{Cumulants of the distribution of $|Q_n|^2$}
\label{s:integrated2}

For sake of brevity, we now drop the subscript $n$ and set $n=1$ until 
the end of this paper, unless otherwide stated. 
All our results can be easily generalized to the 
study of higher order $v_n$'s by multiplying all azimuthal angles by $n$. 

The moments of the $|Q|^2$ distribution involve 
the multiparticle azimuthal correlations discussed in Sec.\ 
\ref{s:multiparticle2}. 
While  $\mean{|Q|^2}$ 
involves two-particle azimuthal correlations, as seen in  
Eq.\ (\ref{qq*}), the higher moments $\mean{|Q|^{2k}}$ 
involve $2k$-particle correlations. For instance, we have 
\begin{equation}
\label{q4dec}
\mean{|Q|^4} =
{1\over M^2}\sum_{j,k,l,m} \mean{e^{i(\phi_j+\phi_k-\phi_l-\phi_m)}}. 
\end{equation} 
These higher order azimuthal correlations can be used to eliminate  
nonflow correlations order by order, as explained in Sec.
\ref{s:multiparticle2}. This will be achieved by taking 
the cumulants of the distribution of $|Q|^2$, 
which we shall soon define.

\subsubsection{Isotropic source}
\label{s:sub3b1}

Following the procedure outlined in Sec.\ \ref{s:multiparticle2}, 
we first consider an isotropic source (no flow). 
Using Eq.\ (\ref{q2order}), $\mean{|Q|^2}$ is then of order unity, 
and so are the higher order moments $\mean{|Q|^{2k}}$. 
However, by analogy with the cumulant decomposition of 
multiparticle distributions introduced in Sec.\ \ref{s:multiparticle}, we can construct 
specific combinations of the moments,
namely the cumulants of the $Q$ distribution, which are much smaller
than unity: the cumulant $\cumul{|Q|^{2k}}$ to order $k$, built 
with the $\mean{|Q|^{2j} }$ where $j \leq k$, 
is of magnitude $1/M^{k-1}$. 

As an illustration, let us construct the fourth order cumulant
$\cumul{|Q|^4}$. 
If the multiplicity $M$ is large, most of the terms in Eq.\ (\ref{q4dec})
are nondiagonal, i.e.\ they correspond to values of $j$, $k$, $l$ 
and $m$ all different. 
Then, using the cumulant of the four-particle azimuthal correlation 
defined by Eq.\ (\ref{defcumul4}) and summing over $(j,k,l,m)$, 
it is natural to define $\cumul{|Q|^4}$ as 
\begin{equation}
\label{eq2p-2p_c} 
\cumul{|Q|^4 } = \mean{|Q|^4 } - 2 \mean{|Q|^2}^2. 
\end{equation}
The order of magnitude of $\cumul{|Q|^4}$  is easy to derive: 
each term of type (\ref{defcumul4}) is of order $1/M^3$ 
as discussed in Sec.\ \ref{s:multiparticle2}; 
there are $M^4$ such terms in the sum (\ref{q4dec}); 
taking into account the factor $1/M^2$ in front of the sum, 
$\cumul{|Q|^4}$ is finally of order $1/M$. 
As intended, two-particle nonflow correlations, which are of 
order unity, have been eliminated in the subtraction (\ref{eq2p-2p_c}). 

A more careful analysis must take into account diagonal terms for which 
two (or more) indices among $(j,k,l,m)$ are equal.  
This analysis is presented in Appendix \ref{s:app-diag}, where we show 
that diagonal terms are also of order $1/M$: they give a
contribution of the same order of magnitude as direct four-particle 
correlations. 
In the following, we shall assume that this property, namely that the 
contribution of diagonal terms is at most of the magnitude of 
the contribution of nondiagonal terms, also holds for higher order moments. 

Among these diagonal terms are the autocorrelations 
already encountered 
in the expansion of $|Q|^2$ [see the discussion below Eq.\ (\ref{q2})], 
which we define as the terms which remain in the absence of flow and 
direct correlations. 
A straightforward calculation (see Appendix \ref{s:app-diag}) shows that 
their contribution to the cumulant 
$\cumul{|Q|^4 }$ is 
$-1/M$.  
As in the case of the second order moment $\mean{|Q|^2}$
discussed previously, autocorrelations are {\it a priori} 
of the same order of magnitude as other nonflow correlations. 
As we shall see later in this section, they can easily be calculated 
and removed order by order. 
\begin{center}
\begin{figure}[ht!]
\centerline{\includegraphics*[width=0.78\linewidth]{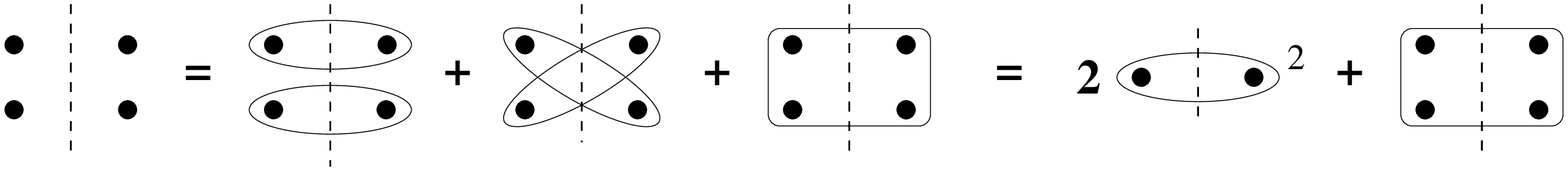}}
\caption{Decomposition of $\mean{|Q|^4} = \mean{QQQ^*Q^*}$. 
In the right-hand side, the first term is of order unity while the 
second term is of order $1/M$.}
\label{fig:fig2p-2p}
\end{figure}
\end{center}

Arbitrary moments $\mean{|Q|^{2k}}$ can be decomposed 
into cumulants, which can then be isolated in a similar way. 
This decomposition 
can be represented in terms of diagrams, like the decomposition 
of the multiparticle distribution in Sec.\ \ref{s:multiparticle2}. 
This is explained in detail in Appendix \ref{s:app-integrated}. 
For example, the decomposition of $\mean{|Q|^4}$ is displayed
in Fig.\ \ref{fig:fig2p-2p}. 
In these diagrams, each dot on the left (resp.\ on the right) 
of the dashed line represents a power of $Q$ (resp.\ $Q^*$),  
and correlated parts, which correspond to direct correlations, 
are circled: the equation displayed in Fig.\ \ref{fig:fig2p-2p}
stands for
\begin{equation}
\mean{|Q|^4}=2\cumul{|Q|^2}^2+\cumul{|Q|^4}. 
\end{equation}
Since $\cumul{|Q|^2}=\mean{|Q|^2}$, one recovers 
Eq.\ (\ref{eq2p-2p_c}). 
More generally, to decompose $\mean{|Q|^{2k}}$, 
one draws $k$ dots on each side of the dashed line. 
The diagrams combine all possible subsets 
of the dots on the left with subsets of the dots on the right 
containing the same number of elements. 
The latter condition is due to the fact that the average value of 
$\mean{Q^l Q^{*m}}$ vanishes when $l\not= m$, as a 
consequence of isotropy. 

In order to invert these relations, and to 
express the cumulants as a function of the 
measured moments, the simplest way consists in using the formalism 
of generating functions, recalled in Appendix \ref{s:b2}. 
There, it is shown that the cumulant $\cumul{|Q|^{2k}}$ 
is obtained from the expansion in power series of $x$ of the 
following generating equation, 
and then the identification of the coefficients of $x^{2k}$:
\begin{eqnarray}
\sum_{k=1}^\infty \frac{x^{2k}}{(k!)^2} 
\cumul{|Q|^{2k}} 
&=& \ln\left(\sum_{k=0}^\infty
\frac{x^{2k}}{(k!)^2} \mean{|Q|^{2k}}\right)\cr
&=& \ln \mean{I_0(2x|Q|)},
\label{geneqi}
\end{eqnarray}
where $I_0$ is the modified Bessel function of order 0. 
Expanding this equation to order $x^4$, one recovers 
Eq.\ (\ref{eq2p-2p_c}); to order $x^6$, one obtains 
the sixth order cumulant
\begin{equation}
\cumul{|Q|^6 } = 
\mean{|Q|^6} - 9 \mean{|Q|^4} \mean{|Q|^2} + 12 \mean{|Q|^2}^3,
\label{eq3p-3p_c}
\end{equation} 
which is of the order of $1/M^2$ for an isotropic 
source.

\subsubsection{Contribution of flow}
\label{s:sub3b2}

Let us now consider small deviations from isotropy. 
As explained in Sec.\ \ref{s:multiparticle2},
these deviations will contribute to the cumulants 
$\cumul{|Q|^{2k}}$ defined above. 

The contribution of flow to the fourth order cumulant 
$\cumul{|Q|^4}$ is calculated in detail in Appendix \ref{s:app-q4}. 
It is shown in particular that the diagonal terms in 
Eq.\ (\ref{q4dec}) are at most of the same magnitude as 
nondiagonal terms, as in the case of an isotropic source. 
As in Sec.\ \ref{s:multiparticle2}, higher order harmonics
can be neglected as soon as condition (\ref{limitv2}) 
is fulfilled. One then obtains 
\begin{equation}
\label{q4}
\cumul{|Q|^4} 
=  -\mean{Q}^4 - {1\over M} + O\left({1\over M}\right)
\end{equation}
where the term $-1/M$ is the contribution of autocorrelations, 
i.e.\ the case $j=k=l=m$. 
From Eqs.\ (\ref{v-Q}) and (\ref{q4}), one can measure values 
of the integrated flow $v$ down to 
$M^{-3/4}$, instead of $M^{-1/2}$ with traditional methods. 

Increased sensitivity can be attained using higher order cumulants. 
As shown in Appendix \ref{s:b3}, the cumulants defined by 
Eq.\ (\ref{geneqi}) are related to the flow by the following 
generating equation:
\begin{equation}
\label{geneq}
\sum_{k=0}^\infty \frac{x^{2k}}{(k!)^2} 
\cumul{|Q|^{2k}} 
= \ln I_0(2x\mean{Q }) +
M\ln I_0\left({2x\over\sqrt{M}}\right).  
\end{equation}
Expanding this equation up to order $x^{2k}$, and isolating the 
coefficient of $x^{2k}/(k!)^2$, one obtains a relation with 
on the left-hand side the cumulant $\cumul{|Q|^{2k} }$, 
while the first term on the right-hand side is the contribution of flow, 
and the second term corresponds to autocorrelations. 
This identity holds within an error of order 
$M^{1-k}$ due to direct $2k$-particle correlations. 
Using Eq.\ (\ref{v-Q}), 
it therefore allows measurements of $v$ within $O(M^{-1+1/2k})$, 
as expected from the discussion of Sec.\ \ref{s:multiparticle2}. 
Expanding Eq.\ (\ref{geneq}) to order $x^4$, 
one recovers (\ref{q4}). To order $x^6$, one obtains 
\begin{equation}
\label{q6}
\cumul{|Q|^6 } 
  =  4\mean{Q}^6 + {4\over M^2} + O\left({1\over M^2}\right)
\end{equation}
which extends the limit of detectability down to $v\sim M^{-5/6}$. 

Since Eqs.\ (\ref{geneqi}) and (\ref{geneq}) 
can be expanded to any order, one obtains 
an infinite set of equations to determine the same quantity 
$\mean{Q}$. The best choice for the order $k$ 
will be discussed below in Sec.\ \ref{s:statistical}. 
Before we come to this point, we shall discuss an alternative method 
to measure $\mean{Q}$, the so-called ``subevent'' method.

\subsection{Subevents}
\label{s:subevents}

The standard flow analysis, instead of studying the 
autocorrelation of the event flow vector as in
Sec.~\ref{s:integrated2}, deals with ``subevents'': 
the set of detected particles is divided randomly into 
two subsets ${\rm I}$ and ${\rm I\!I}$ of equal multiplicities, and 
the two corresponding (subevent) flow vectors $Q_{\rm I}$ and 
$Q_{\rm I\!I}$ are constructed. Then one 
studies the azimuthal correlation between $Q_{\rm I}$ and
$Q_{\rm I\!I}$~\cite{danielewicz85,poskanzer98}. 
This is usually done under the assumption that the only 
azimuthal correlation between the subevents is due to flow. 
Then, from the flow of two equivalent subevents, one can deduce the 
flow of the whole event by a simple multiplication by a factor of
$\sqrt{2}$, as will soon be explained.

A nice feature of that method is that, since the subevents 
have no particle in common, 
autocorrelations are automatically removed: only correlations 
due to flow and direct correlations remain. 
Therefore, one may prefer to work with subevents when direct 
correlations are small (although they are, generally, of the 
same order of magnitude as autocorrelations). 

In this section, we shall improve the standard subevent method, in 
the spirit of Sec.~\ref{s:integrated2}: we eliminate nonflow 
azimuthal correlations order by order by means of a cumulant expansion 
of the distribution of $Q_{\rm I}$ and $Q_{\rm I\!I}$, thereby increasing 
the sensitivity of the method.

\subsubsection{Definitions}

Consider two separate subevents of multiplicity $M_{\rm I}$ and
$M_{\rm I\!I}$ respectively (in practice, one chooses $M_{\rm I}=M_{\rm I\!I}$). 
We can construct the subevent flow vectors 
${Q_{\rm I}}_n$ and ${Q_{\rm I\!I}}_n$ as following:
\begin{eqnarray}
\label{sub-qn}
{Q_{\rm I}}_n &=& \frac{1}{\sqrt{M_{\rm I}}} 
              \sum_{j=1}^{M_{\rm I}} e^{i n \phi_j}
           =|{Q_{\rm I}}_n| e^{i n \Psi_{\rm I}}, \\
{Q_{\rm I\!I}}_n &=& \frac{1}{\sqrt{M_{\rm I\!I}}}
                   \sum_{k=1}^{M_{\rm I\!I}} e^{i n \tilde{\phi}_k}
                = |{Q_{\rm I\!I}}_n| e^{i n \Psi_{\rm I\!I}}.
\end{eqnarray}
As in Sec.\ \ref{s:integrated2}, we set $n=1$ and drop the subscript
$n$ in ${Q_{\rm I}}_n$ and ${Q_{\rm I\!I}}_n$; generalization to
higher $n$ is straightforward.

By analogy with Eq.\ (\ref{v-Q}), we write 
\begin{equation}
\mean{Q_{\rm I}} = \sqrt{M_{\rm I}}\, v_{\rm I}, 
\qquad
\mean{Q_{\rm I\!I}} =\sqrt{M_{\rm I\!I}}\, v_{\rm I\!I},
\label{sub-order}
\end{equation}
where $v_{\rm I}$ and $v_{\rm I\!I}$ denote the values of $v_1$ 
associated with each subevent. 
Hereafter, we shall assume that the two subevents are equivalent, 
i.e.\ $v_{\rm I}=v_{\rm I\!I}\equiv v$, as is the case if they 
are chosen randomly. 

Therefore, if $M_{\rm I}=M_{\rm I\!I}=M/2$, the value of $\mean{Q}$
for the whole event is related to the value for one subevent, 
$\mean{Q_{\rm I}}$, by
\begin{equation}
\label{resol}
\mean{Q} = \sqrt{M} v = \sqrt{2} \, \mean{Q_{\rm I}}.
\end{equation}
The purpose here is to measure $\mean{Q_{\rm I}}$, which is equivalent 
to measuring $\mean{Q}$.

\subsubsection{Limitations of the standard method}

In order to extract the azimuthal correlation between the subevents,
the simplest possibility is to form the product 
\begin{equation}
\label{sub-qq*}
Q_{\rm I} Q_{\rm I\!I}^*={1\over \sqrt{M_{\rm I}M_{\rm I\!I}}} 
\sum_{j,k} e^{i(\phi_j-\tilde{\phi}_k)} = 
|Q_{\rm I} Q_{\rm I\!I}| e^{i(\Psi_{\rm I}-\Psi_{\rm I\!I})}.
\end{equation}
Using Eq.\ (\ref{c2}), the average over many events gives
\begin{equation}
\label{sub-q2}
\mean{Q_{\rm I} Q_{\rm I\!I}^* } = 
\sqrt{M_{\rm I}M_{\rm I\!I}} \left(v^2+
\connex{e^{i(\phi_j-\tilde{\phi}_k)}}\right). 
\end{equation}
This equation is analogous to Eq.\ (\ref{q2}), with the important
difference that autocorrelations 
[the first term in the right-hand side of Eq.\ (\ref{q2})]
no longer appear. 
As a consequence, $\mean{Q_{\rm I}Q_{\rm I\!I}^* }$ vanishes 
if there are no azimuthal correlations between
particles. 

However, two-particle nonflow correlations do remain.
Since they are of order $1/N$, Eq.\ (\ref{sub-q2}) can be written
\begin{equation}
\label{sub-q2order}
\mean{Q_{\rm I} Q_{\rm I\!I}^*} = \sqrt{M_{\rm I}M_{\rm I\!I}} 
\left[v^2 + O\left(\frac{1}{N}\right)\right]. 
\end{equation}
One recognizes in the right-hand side of this equation 
the flow and nonflow contributions to two-particle azimuthal
correlations, as in Eq.\ (\ref{c2order}). 
The only difference lies in the global multiplicative factor
$\sqrt{M_{\rm I}M_{\rm I\!I}}$. 
In particular, summing over many particles does not decrease 
the relative weight of nonflow correlations, as might be believed:
they add up in the same way as the correlations due to flow.

\subsubsection{Beyond the standard method}

Now, following the procedure outlined in Sec.\ \ref{s:integrated2}, 
it is possible to eliminate nonflow correlations between $Q_{\rm I}$ and 
$Q_{\rm I\!I}$ order by order.
This is done by means of a cumulant expansion, which is a trivial 
generalization of the one presented previously. 
The equation to an arbitrary order $2k$ is obtained by replacing, 
in Eqs.\ (\ref{geneqi}) and (\ref{geneq}), $|Q|^2$ with 
$Q_{\rm I}Q_{\rm I\!I}^*$, and 
$\mean{Q}^2$ with $\mean{Q_{\rm I}}\mean{Q_{\rm I\!I}}$.
For example, Eqs.\ (\ref{eq2p-2p_c}) and (\ref{q4}) become 
\begin{equation}
\label{sub-q4}
\mean{Q_{\rm I}^2Q_{\rm I\!I}^{*2} } - 2\mean{Q_{\rm I}Q_{\rm I\!I}^* }^2 = 
M_{\rm I}M_{\rm I\!I} \left[-v^4 + O\left
({1\over N^3}\right)\right],
\end{equation}
which allows measurements of the flow when $v$ is much
larger than $1/N^{3/4}$: the sensitivity is better than with
Eq.\ (\ref{sub-q2order}), where $v$ is to be compared with
$1/N^{1/2}$. The term $-1/M$ in Eq.\ (\ref{q4}), 
which reflects autocorrelations, is automatically removed 
in Eq.\ (\ref{sub-q4}).

To make  a long story short, the same techniques apply to subevents 
as to the whole event. 
The only interest of subevents is that they remove autocorrelations. 
However, they do not remove direct nonflow correlations, which may be 
of the same order of magnitude. Furthermore, autocorrelations can 
be also be subtracted systematically when working with the whole event, 
as shown in Sec.\ \ref{s:integrated2}. 
Another drawback of the subevent method is that each subevent 
contains at most half of the total multiplicity, resulting in 
increased errors. 
As a conclusion, the subevent method seems to be obsolete when working 
with cumulants.

\subsection{Statistical errors}
\label{s:statistical}

The cumulant expansion allows in principle the measurement of $v$ down to 
values of $1/N$ by going to large orders $k$, 
as explained in Sec.\ \ref{s:multiparticle2}. 
In practice, however, since the number of events $N_{\rm evts}$ 
used in the analysis is finite, the sensitivity is limited by statistical 
errors. 
In this section, we determine, as a function of $M$ and $N_{\rm evts}$, 
which order of the cumulant expansion should be chosen so as 
to obtain the most accurate value of the integrated flow. 

First, there is a ``systematic'' error, which is the 
error due to nonflow correlations. 
Expanding Eq.\ (\ref{geneq}) to order $2k$, we obtain an 
equation relating the measured cumulant $\cumul{|Q|^{2k}}$ 
and the integrated flow $\mean{Q}$, which is of the type  
\begin{equation}
\label{arbitraryk}
\cumul{|Q|^{2k}}= a_k\mean{Q}^{2k}+O\left(M^{1-k}\right),
\end{equation}
where $a_k$ is a numerical coefficient of order unity, 
and the last term is the systematic error. 
The resulting error on $\langle Q\rangle$ is therefore
\begin{equation}
\label{ersyst}
\delta\langle Q\rangle_{\rm syst}\sim 
\langle Q\rangle^{1-2k}\,M^{1-k}
\end{equation}
The systematic error thus decreases with increasing $k$, 
since $\langle Q\rangle\sqrt{M}=M v_n\gg 1$, 
as assumed in Eq.\ (\ref{limitv}). 

Let us now discuss the statistical error. 
When averaging a quantity over a large number of events 
$N_{\rm evts}$, the statistical error 
is generally of relative order $1/\sqrt{N_{\text{evts}}}$. 
Since the moments of the distribution of $|Q|^2$ are of order unity, 
the absolute statistical error on the moments is of order 
$1/\sqrt{N_{\text{evts}}}$. The same error applies to the 
cumulants, which are constructed from the moments.
If there is no flow, a more accurate calculation shows 
that the statistical error on the cumulant is 
\begin{equation}
\delta\cumul{|Q|^{2k}}_{\rm stat}={k!\over\sqrt{N_{\rm evts}}}. 
\end{equation}
If the flow is weak, that is if $\langle Q\rangle\ll 1$, this 
formula still holds approximately. Using Eq.\ (\ref{arbitraryk}), 
one thus derives the statistical error on the integrated flow 
\begin{equation}
\label{erstat}
\delta\langle Q\rangle_{\rm stat}\sim 
\langle Q\rangle^{1-2k}\,N_{\rm evts}^{-1/2}.
\end{equation}
Since we have assumed that $\langle Q\rangle \ll 1$, 
the statistical error increases with $k$, unlike the systematic
error. 

It is very likely that this property still holds in the 
more general case when $\langle Q\rangle$ is not much 
smaller than unity. However, we have not been able to derive 
a general formula for the statistical error for arbitrary 
$\langle Q\rangle$ and $k$. 
We only have formulas for the lowest order cumulants. 
Using the cumulant to order 2 ($k=1$), 
\begin{equation}
\delta\langle Q\rangle_{\rm stat}={1\over 2\langle Q\rangle}
\sqrt{1+2\langle Q\rangle^2\over N_{\rm evts}},
\end{equation}
and with the fourth order cumulant ($k=2$),
\begin{equation}
\label{erstat4}
\delta\langle Q\rangle_{\rm stat}={1\over 2 \langle Q\rangle^3}
\sqrt{1+4\langle Q\rangle^2+\langle Q\rangle^4+2\langle Q\rangle^6
\over N_{\rm evts}}.
\end{equation}
One sees on these two formulas that for very strong 
flow ($\langle Q\rangle\gg 1$), the statistical error 
$\delta\langle Q\rangle_{\rm stat}$ is of order 
$1/\sqrt{N_{\rm evts}}$, independent of $\langle Q\rangle$. 
This remains true for higher order cumulants. 
Note, moreover, that both formulas give Eq.\ (\ref{erstat}) in 
the limit of small $\mean{Q}$. 

Since the systematic error decreases with $k$ and the 
statistical error increases with $k$, the best 
accuracy is achieved for the value of $k$ such that 
both are of the same order of magnitude. 
Using Eqs.\ (\ref{ersyst}) and (\ref{erstat}), one thus obtains 
the optimal value of the order $2k$: 
\begin{equation}
2k\simeq  2+\frac{\ln N_{\text{evts}}}{\ln M}. 
\label{err-stat2}
\end{equation}
Since, in practice, $M$ is at least of the order of a hundred at 
ultrarelativistic energies, the fourth order cumulant $2k=4$ 
[i.e.\ Eq.\  (\ref{q4})] gives the best accuracy if the number of 
events lies in the range $10^3<N_{\rm evts}<10^6$. 
Higher order cumulants may be useful if a large statistics is 
available and/or if the multiplicity $M$ is low, as for instance 
in a peripheral collision. 

The flow is detectable only if $\langle Q\rangle$ is larger than 
both statistical and systematic errors. 
Taking for instance $N_{\rm evts}=10^5$ and $M=300$, statistical 
and systematic
errors are of the same order. One then obtains, using Eq.\ (\ref{erstat4}), 
that flow can be seen if $\langle Q\rangle >0.3$. 
Using Eq.\ (\ref{v-Q}), $v$ can be measured down to $1.6\%$ using 
the fourth order cumulant. 
If $v=3\%$, a typical value at the CERN SPS, then 
$\langle Q\rangle \simeq 0.5$. Using Eq.\ (\ref{erstat4}), the 
typical error is then $\delta\langle Q\rangle \simeq 0.02$, 
i.e.\ $\delta v=0.1\%$.

\subsection{Weighted $Q$-vectors}
\label{s:weighted}

The vector $Q_n$ has been defined in Eq.\ (\ref{qn}) with unit weights. 
A more general definition is 
\begin{equation}
Q_n = \frac{1}{\sqrt{\sum_{j=1}^M w^2_j}} \sum_{j=1}^M w_j e^{in\phi_j},
\label{qnw}
\end{equation}
where the weight $w_j$ is an arbitrary function of $p_T$, $y$, the particle 
type, and the order of the harmonic under study. As a consequence, we shall 
restore the index $n$ in this subsection.

\subsubsection{Flow analysis with arbitrary weights}
\label{s:sub3d1}

The method exposed in Sec.\ \ref{s:integrated2} also applies 
with this more general definition. 
There are only two slight differences. 
The first is that the average value of $Q_n$, which we 
have denoted by $\mean{Q_n}$, is no longer related to the 
average value $v_n$ of the flow by Eq.\ (\ref{v-Q}). 
This modification is not important for what follows: 
we shall see in Sec.\ \ref{s:differential2} that measurements
of differential flow depend on the value of $\mean{Q_n}$ rather 
than $v_n$. The second difference is that autocorrelations 
cannot be removed so simply: the procedure given in Appendix 
\ref{s:b4} is no longer valid, so that the subevent method, 
which avoids autocorrelations, may regain some interest. 

Apart from this difference, the procedure is the same as in 
Sec.\ \ref{s:integrated2}. In particular, the cumulants of the 
event flow vector distribution are expressed in the same way in 
terms of the moments. The generating equation (\ref{geneq}) 
still holds, with the caveat that the last term, corresponding 
to autocorrelations, is no longer exact. 

However, autocorrelations are unchanged at the lowest order: 
a calculation analogous to the one leading to 
Eq.\ (\ref{q2}) shows that $\mean{|Q_n|^2}=1$ if there 
are no azimuthal correlations between particles, up to terms of order 
$1/M$. 
Changes occur only at higher orders.

\subsubsection{Optimal weights}
\label{s:sub3d2}

What is the best choice for the weight $w(p_T,y,n)$?
In practice, it should be chosen so as to maximize 
the effect of flow: one should try to obtain a value of 
$\mean{Q_n}$ as large as possible, since this value 
will determine the accuracy in the measurement 
of azimuthal distributions, as we shall see in 
Sec.\ \ref{s:differential}. 
From the definition (\ref{qnw}), averaging over azimuthal angles 
and denoting by $(v_n)_j$ the value of $v_n$ for the corresponding 
particle, one obtains 
\begin{equation}
\mean{Q_n} = {\sum_{j=1}^M  (v_n)_j w_j
\over\sqrt{\sum_{j=1}^M w^2_j}}\le 
\sqrt{\sum_{j=1}^M (v_n)^2_j},
\end{equation}
where we have used a simple triangular inequality, and the fact that 
the flow coefficients $(v_n)_j$ are real. 
The identity holds when $w_j=\lambda (v_n)_j$, 
where $\lambda$ is arbitrary. 
In other terms, the optimal weight for a particle with 
given rapidity and transverse momentum is the associated flow 
coefficient $(v_n)_j$ itself. 

Of course, since the goal is precisely to measure $v_n$, the 
above discussion does not answer the question of the choice of the 
optimal weight. 
However, general properties of the $v_n$'s 
can be used to guess a reasonable choice of $w$. 
Since $v_n$ is an odd (resp.\ even) function of the center of mass 
rapidity for odd $n$ (resp.\ even $n$), so should be $w$. 
Regarding the $p_T$ dependence, one may note that at low 
$p_T$, $v_n$ generally behaves as $v_n\propto p_T^n$
\cite{ollitrault98}. Therefore, it seems natural to choose 
$w\propto p_T^n$ when measuring the $n^{\rm th}$ harmonic. 
For $n=1$, $Q_n$ then becomes the sum of transverse momenta, 
weighted by an odd function of rapidity, which was the definition 
chosen in \cite{danielewicz85}.
For $n=2$, $Q_n$ is then equivalent to the transverse momentum 
sphericity tensor used in \cite{ollitrault92}.

\subsection{Gaussian limit} 
\label{s:gaussian}

In this section, we compare our method to methods previously 
used in \cite{voloshin96,ollitrault92}, which rely on the 
large multiplicity, Gaussian limit. 
It is well known that, according to the central limit 
theorem, the distribution of the fluctuations of 
$Q$ around its average value $\mean{Q }$ is Gaussian in the 
limit of large $M$. 
Up to corrections of order $1/M$,
the normalized probability of $Q=Q_x+iQ_y$ thus reads
\begin{equation}
\frac{{\rm d}p}{{\rm d}^2Q} = 
\frac{1}{2\pi\sigma_x\sigma_y} \exp\left(
-\frac{(Q_x - \mean{Q})^2}{2\sigma_x^2}
-\frac{Q_y^2}{2\sigma_y^2}
\right),
\label{pQn1}
\end{equation}
with 
$\sigma_x^2 = \mean{Q_x^2 } - \mean{Q }^2$ and 
$\sigma_y^2 = \mean{Q_y^2 }$.

We shall first show that this limit
is equivalent to the cumulant expansion to order 4 presented in Sec.\ 
\ref{s:integrated2}. 
Then we shall discuss the relationship with an alternative method to 
measure flow, which has been used in the literature, and consists in 
fitting the distribution of $|Q|$.

\subsubsection{Higher harmonics}

In the case of the Gaussian distribution (\ref{pQn1}), one easily calculates 
the cumulants used in Sec.\ \ref{s:integrated2}:
\begin{eqnarray}
\label{q24gaussian}
\mean{|Q|^2}&=&\mean{Q}^2+\sigma_x^2+\sigma_y^2,\cr
\mean{|Q|^4}-2\mean{|Q|^2}^2&=&-\mean{Q}^4
+2(\sigma_x^2-\sigma_y^2)\mean{Q}^2
+(\sigma_x^2-\sigma_y^2)^2.
\end{eqnarray}
In order to compare these equations with Eqs.\ (\ref{q2order}) and 
(\ref{app-q4order}), 
we need to evaluate the sum $\sigma^2\equiv\sigma_x^2+\sigma_y^2$ 
and the difference $\sigma_x^2-\sigma_y^2$.

From Eqs.\ (\ref{v-Q}) and (\ref{q2}), one obtains 
\begin{eqnarray}
\sigma^2&=& \mean{|Q|^2 } - \mean{Q }^2\cr
&=&1-v_1^2+(M-1)\connex{e^{i(\phi_j-\phi_k)}}, 
\label{sig2}
\end{eqnarray}
where the last term is of order unity since 
$\connex{e^{i(\phi_j-\phi_k)}}$ is of order $1/N\lesssim 1/M$. 
This still holds for the generalized vector (\ref{qnw}). 
One thus recovers Eq.\ (\ref{q2order}). 

Let us now calculate the difference:
\begin{eqnarray}
\sigma_x^2-\sigma_y^2&=&{1\over M}\sum_{j,k=1}^M\left[
\mean{\cos(\phi_j+\phi_k)}-\mean{\cos\phi_j} \mean{\cos\phi_k}\right]\cr
&=&v_2-v_1^2+O(v_2),
\end{eqnarray}
where the first two terms in the last equation come from the 
diagonal terms $j=k$, while the remaining term is the 
contribution of nondiagonal terms. 
Reporting this expression into Eq.\ (\ref{q24gaussian}), 
we recover Eq.\ (\ref{app-q4order}): higher harmonics reflect 
a deviation from isotropy in the fluctuations of $Q$.

\subsubsection{Isotropic fluctuations}
\label{s:sub3c1}

Neglecting higher harmonics, we may write $\sigma_x=\sigma_y$. 
Then the distribution (\ref{pQn1}) becomes 
\begin{equation}
\frac{{\rm d}p}{{\rm d}^2Q} = 
\frac{1}{\pi\sigma^2} \exp\left(-\frac{|Q - \mean{Q}|^2}{\sigma^2}\right). 
\label{pQniso}
\end{equation}
With this distribution, we expect to recover the results 
of Sec.\ \ref{s:integrated2}, where higher harmonics were
also neglected. 

Indeed, one finds after some algebra, 
for arbitrary, real $x$,
\begin{equation}
\label{geneqgaussian}
\ln \mean{I_0(2x|Q|) } = \sigma^2x^2 +\ln I_0(2x\mean{Q }),
\end{equation}
to be compared with Eqs.\ (\ref{geneqi}) and (\ref{geneq}). 
According to Eq.\ (\ref{sig2}), 
the extra term $\sigma^2 x^2$ is of order unity, 
in agreement with the statement following 
Eq.\ (\ref{geneq}) that the correction at order $x^{2k}$ is $O(M^{1-k})$. 

Corrections to the central limit theorem are of order $1/M$. 
Thus, expanding Eq.\ (\ref{geneqgaussian}) in powers of $x$, one 
obtains identities which are valid up to that order. 
To order $x^4$, we recover the result obtained in the 
previous section, see Eq.\ (\ref{q4}), with the same accuracy. 
To order $x^{2k}$ with $k>2$, the results obtained in Sec.\ 
\ref{s:integrated2} are more accurate since we have seen that the 
correction is of magnitude $M^{1-k}\ll 1/M$.

\subsubsection{Distribution of $|Q|$}
\label{s:sub3c2} 

A method for extracting the flow from the data, which was proposed 
in \cite{voloshin96,ollitrault92}, consists in plotting 
the measured distribution of $|Q|$. 
This method led to the first observation of collective flow in 
ultrarelativistic nucleus-nucleus collisions \cite{barrette94}. 
It is a simplified version of the method based on the 
sphericity tensor \cite{danielewicz83}, which led to the 
first observation of collective flow at Bevalac \cite{gustafsson84}. 
Note that these methods are more reliable than what we call the 
``standard method'' in this paper, in the sense that one need not 
neglect  nonflow correlations. 

The distribution of $|Q|$ is obtained by integration of Eq.\ (\ref{pQniso}) 
over the phase of $Q$:
\begin{equation}
\frac{1}{|Q|} \frac{{\rm d}p}{{\rm d}|Q|} = 
\frac{2}{\sigma^2} \exp\left(-\frac{\mean{Q}^2+|Q|^2}
{\sigma^2}\right) 
I_0 \left( \frac{2|Q|\mean{Q} }{\sigma^2} \right).
\label{pQn2}
\end{equation}
One must then fit both parameters $\sigma$ 
and $\mean{Q}$ to the data. 

If there is no flow, that is $\mean{Q}=0$, the $|Q|$ distribution 
given by Eq.\ (\ref{pQn2}) is purely Gaussian: 
\begin{equation}
\frac{1}{|Q|} \frac{{\rm d}p}{{\rm d}|Q|} = 
\frac{2}{\sigma^2} \exp\left(-\frac{|Q|^2}{\sigma^2}\right). 
\label{puregaussian}
\end{equation}
The $|Q|$ distribution deviates 
from the Gaussian shape if the flow is strong enough compared 
to the fluctuation scale, 
that is for values of $\mean{Q} \gtrsim \sigma$.
In particular, the maximum of the distribution is shifted to 
$|Q|\not= 0$ if $\mean{Q} > \sigma$. 
Since $\sigma$ is of order 1, using Eq.\ (\ref{v-Q}), 
this condition is equivalent to $v\gtrsim 1/\sqrt{M}$. 
Note, however, that one need not assume $v\gg 1/\sqrt{M}$, 
as with the methods based on two-particle azimuthal correlations. 

If $\mean{Q}\ll\sigma$, i.e.\ $v\ll 1/\sqrt{M}$, the shape of 
the distribution is very close to a pure Gaussian distribution. In fact, the 
deviations from the Gaussian shape are of order 
$\mean{Q}^4/\sigma^4$\cite{ollitrault95}. 
This can be seen by expanding 
Eq.\ (\ref{pQn2}) to order $\mean{Q} ^2$, which  is 
equivalent to replacing $\sigma^2$ with $\sigma^2+\mean{Q} ^2$\
in Eq.\ (\ref{puregaussian}). 
Alternatively, one can eliminate $\sigma$ and obtain 
$\mean{Q}$ directly 
using the following identity, which can be easily
derived from Eq.\ (\ref{pQniso}): 
\begin{equation}
\label{q4gaussian}
\mean{|Q|^4 } - 2\mean{|Q|^2 }^2 = -\mean{Q}^4,
\end{equation}
again showing that the deviation is of fourth order in the flow. 
Knowing that the deviation to the central limit is of order 
$1/M$, one finds that Eq.\ (\ref{q4gaussian}) is equivalent to 
Eqs.\  (\ref{eq2p-2p_c}) and (\ref{q4}), i.e., 
to the cumulant expansion to order 4. 

The importance of the factor $1/\sqrt{M}$ in the definition of 
$Q$, Eq.\ (\ref{qn}), also appears clearly when fitting Eq.\ (\ref{pQn2})
to experimental data. 
Because of this factor, $\sigma$ does not depend on the 
multiplicity $M$ in the limit of large $M$, as discussed above. 
This is especially important when the fit is done using 
events with different multiplicities $M$. 
If there is no flow, the distribution of $|Q|$ is Gaussian with width 
$\sigma$. If $\sigma$ depended on $M$, the distribution would rather 
be a superposition of Gaussian distributions with different widths. 
In this case, the left-hand side of Eq.\ (\ref{q4gaussian}) would be
positive, hiding a possible weak flow. 
This phenomenon probably explains why the first analysis of the 
E877 Collaboration \cite{barrette94} gives zero values of the flow 
in some centrality bins. 

When fitting Eq.\ (\ref{pQn2}) to the data, it is important 
to fit independently $\mean{Q}$, which reflects the flow, 
and $\sigma$, which also involves two-particle correlations, 
according to Eq.\ (\ref{sig2}). 
Assuming that $\sigma$ is the same for all Fourier harmonics, as was 
done by E877~\cite{barrette94}, amounts to neglecting two-particle 
correlations. 

Finally, note that the Gaussian limit can also be applied 
to the subevent method, yielding interesting results: 
in particular, the distribution of the relative angle 
between $Q_{\rm I}$ and $Q_{\rm I\!I}$ is not the 
same for direct correlations and correlations due to flow 
\cite{ollitrault95}.

\section{Differential flow}
\label{s:differential}

In this section, we explain how it is possible to perform detailed 
measurements of azimuthal distributions: typically, one wishes to measure 
$v_n$ for a given type of particle as a function of the rapidity 
$y$ and the transverse momentum $p_T$. In the following, we shall 
call this particle a ``proton,'' but it can be anything else. 
We denote by $\psi$ its azimuthal angle, 
and by $v'_m$ the corresponding differential flow coefficients
$v'_m=\mean{e^{im\psi}}$. 
Unlike the standard method, as stated before, we do not make the assumption 
that all azimuthal correlations are due to flow. 
As in the case of the integrated flow studied in Sec.\ \ref{s:integrated}, 
we get rid of nonflow correlations order by order, by means of 
a cumulant expansion. 

The principle of the method is explained in Sec.\ \ref{s:differential1}. 
In Sec.\  \ref{s:differential2}, 
we show that $v'_m$ can be obtained from the azimuthal correlation 
between $\psi$ and the flow vector $Q$. As in the case of 
integrated flow, the order to which nonflow correlations 
must be eliminated depends in practice on the number of 
events available: this is explained in Sec.\ \ref{s:differential3},
where we also estimate the resulting accuracy on $v'_m$. 
Our method is compared to traditional methods in Sec.\  
\ref{s:differential4}.

\subsection{Principle and orders of magnitude}
\label{s:differential1}

The differential flow coefficients $v'_m$ can be obtained 
only through azimuthal correlations with other particles, 
typically particles used to estimate the orientation of the 
reaction plane, which we call ``pions'' in this section, although 
they can be anything else. 

For instance, correlating the proton with one pion, $v'_1$ can be 
derived from the measurement of the two-particle azimuthal correlation
\begin{equation}
\label{c2diff}
\mean{e^{i(\psi-\phi_1)}} = v'_1 v_1 +O\left({1\over N}\right),
\end{equation}
where $v_1$ refers to the pion, and is determined independently. 
We have used an analogy with Eq.\ (\ref{c2order}). 
The term $O(1/N)$ comes from two-particle nonflow correlations 
between the proton and the pion. 
The error made in the determination of $v'_1$ is thus of order 
$1/(Nv_1)$. Of course, one should correlate the proton to 
particles with a strong flow, so that $v_1$ be as large as possible. 

More accurate measurements can be obtained using higher order 
correlations and a cumulant expansion. 
For instance, at fourth order, 
one can eliminate the two-particle 
nonflow correlation by correlating the proton with 
three pions and taking the cumulant, 
by analogy with Eqs.\ (\ref{defcumul4}) and (\ref{c4order}): 
\begin{eqnarray}
\cumul{e^{i(\psi+\phi_1-\phi_2-\phi_3)}} & \equiv & 
\mean{e^{i(\psi+\phi_1-\phi_2-\phi_3)}} -
\mean{e^{i(\psi-\phi_2)}} \mean{e^{i(\phi_1-\phi_3)}} -
\mean{e^{i(\psi-\phi_3)}} \mean{e^{i(\phi_1-\phi_2)}}\cr 
& = & -v'_1v_1^3+O\left({1\over N^3}\right).
\label{c4diff}
\end{eqnarray}
More generally, correlating the proton with $2k+1$ pions, 
the connected part of the correlation is of order $1/N^{2k+1}$ 
[since it corresponds to direct $(2k+2)$-particle correlations], 
while the contribution of flow is $v'_1 v_1^{2k+1}$. 
Comparing both terms, the accuracy on $v'_1$ is thus of 
order $1/(Nv_1)^{2k+1}$. 
Using Eq.\ (\ref{limitv}), this shows that 
the accuracy increases with increasing $k$, i.e.\ when using 
multiparticle correlations. 

Higher harmonics, such as $v'_2$, can be obtained by at least 
two methods. The first consists in multiplying all the angles by 2 in 
the equations above, and replacing $v'_1$ and $v_1$ with $v'_2$
and $v_2$, respectively. 
A second method is to mix two different harmonics, measuring 
$\mean{e^{i(2\psi-\phi_1-\phi_2)}}$. 
If the source is isotropic, this quantity is of order $1/N^2$ since it 
involves a direct three-particle correlation. If there is flow, 
neglecting other sources of correlation for simplicity, 
$\mean{e^{i(2\psi-\phi_1-\phi_2)}}$ factorizes into 
$\mean{e^{2i\psi}} \mean{e^{-i\phi_1}}
\mean{e^{-i\phi_2}} =v'_2 v_1^2$. 
Putting everything together, we obtain 
\begin{equation}
\label{v21}
\mean{e^{i(2\psi-\phi_1-\phi_2)}} = v'_2 v_1^2 
+O\left({1\over N^2}\right). 
\end{equation}
One sees that nonflow correlations come into play only at order $1/N^2$, 
rather than $1/N$ when comparing the same harmonics as in
(\ref{c2diff}).
Nonetheless, they do not disappear. These correlations can also 
be eliminated order by order using the cumulant expansion, as 
we shall see in Sec.\ \ref{s:differential2}. Generally, 
if one correlates the proton with $2k+m$ pions, one obtains 
an accuracy on $v'_m$ of order $1/(Nv_1)^{2k+m}$.

\subsection{Differential flow from correlations with $Q_n$}
\label{s:differential2}

In order to correlate a proton with pions, it is 
convenient to use the event flow vector $Q_n$, Eq.\ (\ref{qn}). 
From now on in this section, we choose $n=1$, and drop the 
subscript $n$, i.e.\ we write $Q$ and $v$ instead of $Q_1$ and 
$v_1$. On the other hand, we keep the subscript $m$ for the 
proton $v'_m$ because several harmonics may be measured. 
Generalization to arbitrary $n$ is straightforward: 
one simply multiplies all azimuthal angles 
(of both protons and pions) by $n$. 

In the standard flow analysis, one usually excludes
``autocorrelations'' by excluding the ``proton'' under study 
from the definition of the event flow vector \cite{danielewicz85}; 
that is, the azimuthal 
angle $\psi$ is not one of the $\phi_j$ in Eq.\ (\ref{qn}). 
Within our method, one can still do so, but it is not even necessary.
First, autocorrelations 
will be removed order by order as well as direct correlations, 
as in the case of the integrated flow in Sec.\ \ref{s:integrated2}. 
Furthermore, autocorrelations, if any, can be subtracted 
exactly if the event flow vector $Q$ is defined with unit weight, 
as in Eq.\ (\ref{qn}). This subtraction is performed in Appendix \ref{s:c4}. 
For simplicity, we neglect the corresponding term in this section, 
unless otherwise specified. 

Let us start with the measurement of the first harmonic $v'_1$. 
The two-particle azimuthal correlation between the proton and a pion, 
Eq.\ (\ref{c2diff}), can be expressed introducing the vector $Q$
defined by Eq.\ (\ref{qn}). 
Summing Eq.\ (\ref{c2diff}) over all the pions involved in 
$Q$, one obtains the correlation between $\mean{Q}$ and the 
proton: 
\begin{equation}
\label{q2diff}
\mean{Q^* e^{i\psi}} =\mean{Q } 
\left[v'_1+O\left({1\over Nv}\right)\right].
\end{equation}
The value of $\mean{Q}$ must be obtained independently, 
using the methods discussed in Sec.\ \ref{s:integrated}.

More accurate measurements, involving correlations of the 
proton with several pions, are performed using higher order moments, 
as in Sec.\ \ref{s:integrated2}. These higher order moments 
are obtained by weighting the previous expression with powers of $|Q|^2$, 
i.e.\ by measuring $\mean{|Q|^{2k} Q^* e^{i\psi}}$.
These moments are then decomposed into cumulants. 
For instance, Eq.\ (\ref{c4diff}) becomes 
\begin{eqnarray}
\label{q4diff}
\cumul{|Q|^2 Q^* e^{i\psi}} &\equiv& 
\mean{|Q|^2 Q^* e^{i\psi} } 
- 2 \mean{Q^* e^{i\psi}} \mean{|Q|^2 } \cr
&=& - \mean{Q }^3 
\left[v'_1+O\left({1\over (Nv)^3}\right)\right],
\end{eqnarray}
through which we define the cumulant $\cumul{|Q|^2 Q^* e^{i\psi}}$. 

\begin{center} 
\begin{figure}[ht!]
\centerline{\includegraphics*[width=0.43\linewidth]{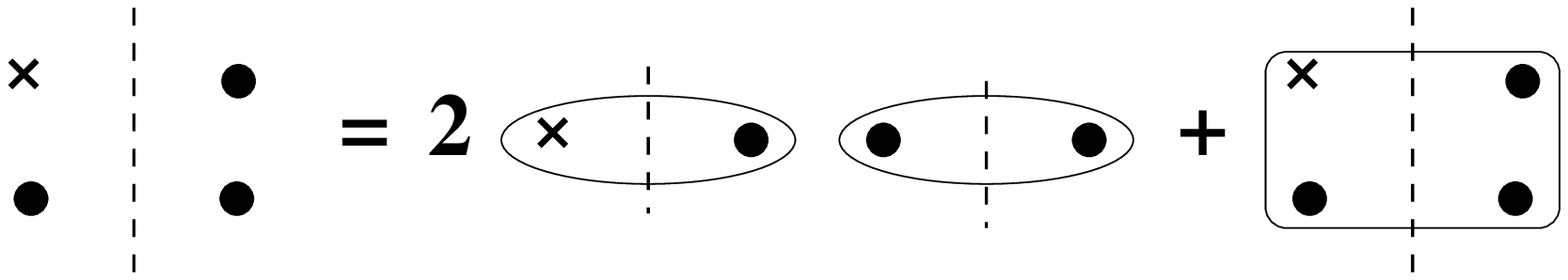}}
\caption{Decomposition of 
$\mean{|Q|^2 Q^* e^{i\psi}}=\mean{Q^{* 2} Q e^{i\psi} }$. 
The cross correspond to the factor $e^{i\psi}$ of the proton, while dots on left 
(resp.\ on the right) of the dashed line stand for $Q$ (resp.\ $Q^*$).}
\protect\label{fig:fig1x1p-2p}
\end{figure}
\end{center}

As in the case of integrated flow, the decomposition 
of higher order moments $\mean{|Q|^{2k}Q^* e^{i\psi}}$ 
in cumulants can be represented in terms of diagrams. 
For instance, the decomposition of $\mean{|Q|^2 Q^* e^{i\psi}}$ 
is displayed in Fig.\ \ref{fig:fig1x1p-2p}. 
The diagrams in this figure stand for 
\begin{eqnarray}
\mean{|Q|^2 Q^* e^{i\psi} } & = &
2 \cumul{Q^* e^{i\psi} } \cumul{|Q|^2 }
+ \cumul{|Q|^2 Q^* e^{i\psi} } \cr 
 & = & 2 \mean{Q^* e^{i\psi} } \mean{|Q|^2 }
+ \cumul{|Q|^2 Q^* e^{i\psi} } 
\label{eq1x1p-2p}
\end{eqnarray}
One thus recovers the expression of the cumulant, Eq.\ (\ref{q4diff}).
More generally, in order to decompose the moment  
$\mean{|Q|^{2k} Q^* e^{i\psi}} = \mean{Q^k Q^{* k+1} e^{i\psi}}$,
one draws a cross on the left representing the proton, 
$k$ dots on the left and $k+1$ dots on the right representing the pions. 
The graphs combine all possible subsets of the points on the left 
with subsets of the points on the right containing the same number 
of elements. 

\begin{center}
\begin{figure}[ht!]
\centerline{\includegraphics*[width=0.44\linewidth]{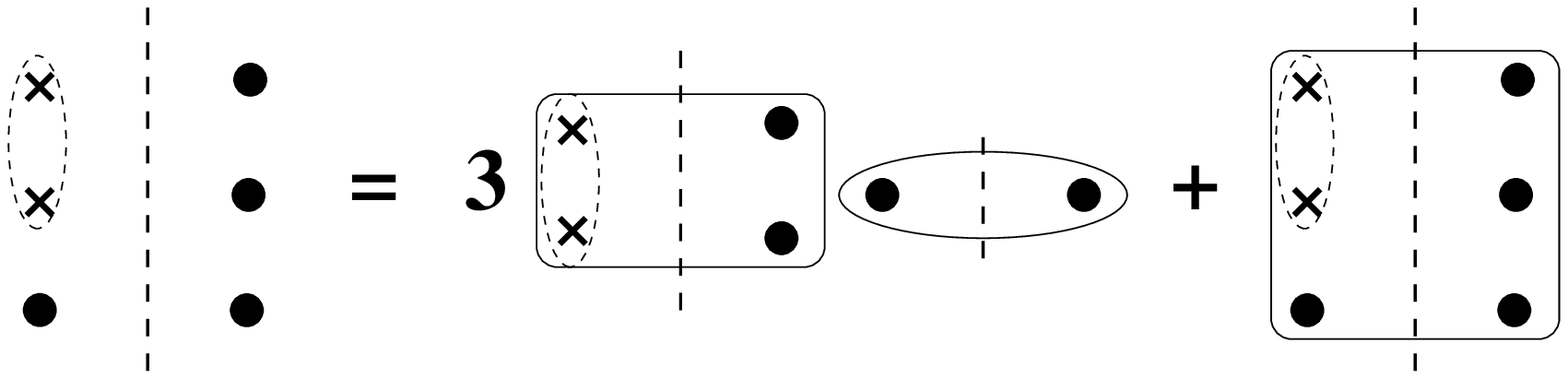}}
\caption{Expansion of $\mean{|Q|^2 Q^{* 2} e^{2i\psi}} =
\mean{Q^{* 3} Q e^{2i\psi}}$. 
The linked crosses stand for the proton, while dots on left (resp.\ on the right) 
correspond to $Q$ (resp.\ $Q^*$).}
\protect\label{fig:fig2x1p-3p}
\end{figure}
\end{center}

Let us now discuss the measurements of higher harmonics of the 
proton azimuthal distribution $v'_m$. 
In the case $m=2$, Eq.\ (\ref{v21}) gives, summing over the pions 
involved in $Q$, 
\begin{equation}
\label{v21bis}
\mean{Q^{* 2} e^{2i\psi}} =\mean{Q }^2 
\left[v'_2+O\left({1\over (Nv)^2}\right)\right].
\end{equation}
To obtain a better accuracy, one must decompose higher order moments 
$\mean{|Q|^{2k}Q^{* 2} e^{2i\psi}}$  in cumulants. 
In terms of the diagrammatic representation, the proton 
is now associated with two crosses, as seen in 
Fig.\ \ref{fig:fig2x1p-3p} for $k=1$.  
As before, the 
graphs combine all possible subsets of the points on the left 
with subsets of the points on the right containing the same number 
of elements, with the subsidiary condition that the two crosses 
belong to the same subset. 
In the left-hand side of Fig.\ \ref{fig:fig2x1p-3p}, 
the dot on the left of the dashed line can be associated with 
any of the three dots on the right. The equation represented by the 
figure can be written 
\begin{eqnarray}
\label{q22diff}
\mean{|Q|^2 Q^{* 2} e^{2i\psi}} & = & 
3\cumul{Q^{* 2} e^{2i\psi}} \cumul{|Q|^2 } 
+ \cumul{|Q|^2 Q^{* 2} e^{2i\psi}} \cr
 & = & 3\mean{Q^{* 2} e^{2i\psi}} \mean{|Q|^2 } 
+ \cumul{|Q|^2 Q^{* 2} e^{2i\psi}}
\end{eqnarray}
where the last term involves a direct five-particle correlation, and is 
therefore of order $M^2\times O(1/N^4)$. 
When there is flow, one obtains
\begin{eqnarray}
\label{q22diffflow}
\cumul{|Q|^2 Q^{* 2} e^{2i\psi}} & = & 
\mean{|Q|^2 Q^{* 2} e^{2i\psi}} - 
3\mean{Q^{* 2} e^{2i\psi}} \mean{|Q|^2 } \cr
 & = & \mean{Q }^4 
\left[-2v'_2+O\left({1\over (Nv)^4}\right)\right].
\end{eqnarray}

Cumulants of arbitrary order, for arbitrary harmonics $v'_m$,  
can be obtained by expanding in powers of $x$ the 
following generating equation, derived in 
Appendix \ref{s:c2}:
\begin{equation}
\sum_{k=0}^\infty \frac{x^{2k+m}}{k!\,(k+m)!} 
\cumul{|Q|^{2k} {Q^*}^m e^{im\psi}} 
=\frac{\mean{I_m(2x|Q|)\,(Q^*/|Q|)^m e^{im\psi} }}{\mean{I_0(2x|Q|)
}} 
\label{gendiffcumul}
\end{equation}
where $I_m$ is the modified Bessel function of order $m$. 
For $m=1$, one recovers Eq.\ (\ref{eq1x1p-2p}) by expanding 
this equation to order $x^3$. 
For $m=2$, one recovers Eq.\ (\ref{q22diff}) by expanding this 
equation to order $x^4$. 

The cumulants defined by Eq.\ (\ref{gendiffcumul}) 
are related to the differential flow by 
\begin{equation}
\sum_{k=0}^\infty \frac{x^{2k+m}}{k!\,(k+m)!} 
\cumul{|Q|^{2k} {Q^*}^m e^{im\psi}} = 
\frac{I_m(2x\mean{Q })}{I_0(2x\mean{Q })}\, v'_m
+{I_m\left({2x/\sqrt{M}}\right)\over
I_0\left({2x/\sqrt{M}}\right)}
\label{geneqdiff}
\end{equation}
The second term corresponds to autocorrelations, and must be included 
only if the proton is involved in the flow vector $Q$.  
In the case $m=1$, one recovers the lowest order formulas (\ref{q2diff}) and
(\ref{q4diff}) by expanding this equation to order $x$ and 
$x^3$, respectively. 
For $m=2$, one recovers Eqs.\ (\ref{v21bis}) and (\ref{q22diffflow}) by 
expanding it to order $x^2$ and $x^4$, respectively. 

At order $x^{2k+m}$, Eq.\ (\ref{geneqdiff})
gives an accuracy on $v'_m$ of order $1/(Nv)^{2k+m}$,
as expected from the discussion of Sec.\ \ref{s:differential1}.

\subsection{Statistical errors}
\label{s:differential3}

Equation (\ref{geneqdiff}) generates 
an infinite set of equations to measure the differential flow 
$v'_m$, since it can be expanded to any arbitrary order $x^{2k+m}$. 
As in the case of integrated flow, the best choice of $k$ is 
the one that yields the best accuracy on $v'_m$. It results from 
a compromise between systematic errors stemming from nonflow correlations, 
which decrease when using higher order cumulants, 
and statistical errors, which increase with the order $k$. 

The equation obtained when expanding Eq.\ (\ref{geneqdiff}) 
to order $x^{2k}$ is of the type 
\begin{equation}
\label{difforder}
\cumul{|Q|^{2k}Q^{* m}e^{im\psi}}= b_k\mean{Q}^{2k+m}
v'_m+O\left(M^{-k-(m/ 2)}\right),
\end{equation}
where $b_k$ is a numerical coefficient of order unity. 
Neglecting for the moment the error on the integrated flow $\mean{Q}$, 
this equation gives a systematic error on $v'_m$
\begin{equation}
\label{erdifsyst}
(\delta v'_m)_{\rm syst}\sim \mean{Q}^{-2k-m} M^{-k-(m/ 2)}.
\end{equation}
This systematic error decreases when increasing the order $k$. 
Note that for $m=1$, the systematic error should be the 
same on the differential flow as on its integrated value
at a given order. Thus we expect Eqs.\ (\ref{erdifsyst}) 
and (\ref{ersyst}) to give the same result, using (\ref{v-Q}). 
In doing the comparison, one must pay attention to the fact 
that the cumulant used for differential flow 
$\cumul{|Q|^{2k}Q^{*}e^{i\psi}}$ involves $2k+2$ 
particles while the cumulant for integrated flow 
$\cumul{|Q|^{2k}}$ involves only $2k$ particles.
Thus, comparing the two ``at a given order'' means that 
we must replace $2k$ in Eq.\ (\ref{erdifsyst}) by $2k+2$ 
in Eq.\ (\ref{ersyst}).

The statistical error on the cumulant 
(\ref{difforder}) is of order $1/\sqrt{N'_{\rm evts}}$, where 
$N'_{\rm evts}$ is the number of events containing a proton. 
This leads to an error on $v'_m$:
\begin{equation}
\label{erdifstat}
(\delta v'_m)_{\rm stat}\sim 
\mean{Q}^{-2k-m}
\,\left(N'_{\rm evts}\right)^{-1/2}
\end{equation}
If $m=1$, we again recover the result obtained for the integrated 
flow, provided we replace $2k$ by $2k+2$ in Eq.\ (\ref{erstat}), and 
$N'_{\rm evts}$ by $MN_{\rm evts}$ (which is the total number of 
particles involved in the measurement of integrated flow) 
in Eq.\ (\ref{erdifstat}). 

If  $\mean{Q} <1$, the statistical error (\ref{erdifstat}) increases with 
increasing $k$, and the optimal value of  $k$ is that for which 
statistical and systematic errors are equivalent, i.e.\ 
\begin{equation}
\label{kdiff}
2k\simeq  -m + \frac{\ln N'_{\text{evts}}}{\ln M}. 
\end{equation}
The error on $\mean{Q}$, estimated in Sec.\ \ref{s:statistical},
should also be taken into account. However, the measurement of 
differential flow is done in a limited region of phase space, 
by definition, so that the corresponding 
statistics is smaller than for the integrated flow where many 
more events can be used. 
It is then safe to assume that the statistical error on $\mean{Q}$ gives a 
negligible contribution to the error on $v'_m$. 

If $m=1$, the previous equation shows that $k=1$ is more accurate than  
$k=0$ (the latter value corresponds to the standard method, 
neglecting correlations)  only if $\ln N'_{\rm evts}/\ln M>2$, i.e.\ 
if the statistics is large enough, typically $N'_{\rm evts}>10^4$
for an event multiplicity $M\sim 100$. 
For higher harmonics $m>1$, the contribution of 
nonflow correlations are
smaller as explained above: thus the lowest order method $k=0$ is 
to be chosen unless a very large number of events is available, typically 
$N_{\rm evts}>10^6$ for the second harmonic $m=2$ if $M\sim 100$.

\subsection{Relation with previous methods}
\label{s:differential4}

Previously used methods\cite{poskanzer98,doss87} 
also study the correlation between the 
event flow vector (\ref{qn}) with the momentum of the proton. 
The traditional justification is that, 
as explained in Sec.\ \ref{s:qvector}, the phase $n\phi_Q$ of 
the event flow vector (\ref{qn}) gives an estimate of the 
orientation of the reaction plane modulo $2\pi/n$. 
Studying the correlation between $\psi$ and $\phi_Q$, 
one can reconstruct harmonics $v'_n$, $v'_{2n}$, $v'_{3n}$, etc. 

The standard analysis relies on a purely angular correlation. 
One measures the average
$\mean{\cos m(\psi-\phi_Q)}$. 
Neglecting nonflow correlations, this quantity is the product of 
$v'_m$ and a resolution factor 
which is given by an independent measurement\cite{ollitrault97}. 
Our method relies on similar averages, weighted by powers of
$|Q|$:
\begin{equation}
\mean{|Q|^{2k+m}\cos m(\psi-\phi_Q)}
=\mean{Q^{* m+k}Q^k e^{im\psi}} .
\end{equation}

In the traditional method, autocorrelations are usually removed explicitly 
by specifying that the proton under study is not used in 
constructing $Q_n$ in Eq.\ (\ref{qn}) \cite{danielewicz85}. 
However, nonflow direct correlations, which are of the same order of 
magnitude as autocorrelations, do remain, and limit the sensitivity of the 
analysis. 
With our method, autocorrelations can be 
removed in the same way as in the standard analysis. 
But we also remove direct correlations, thereby increasing the 
sensitivity of the measurements.

\section{Acceptance corrections}
\label{s:acceptance}

For simplicity, the discussion has been limited so far to an ideal 
detector, i.e.\ a detector with an acceptance which is azimuthally 
isotropic in $\phi$. An actual detector is never perfect, either 
because its components are of uneven quality, or simply because 
it does not cover the whole $\phi$ range. 
In this section, we discuss a simple extension of the method 
which allows us to work with {\it any\/} detector. 
More precisely, it allows the detection of deviations from an isotropic 
source, i.e.\ flow, with any detector, and the correction is
implemented in the same way for all detectors. However, the accuracy 
on the measurement of $v_n$ can be poor if the detector covers 
only a limited range in $\phi$. 

The only modification lies in the definition of the cumulants,
for which the expressions given in Secs.\ \ref{s:integrated2} 
and \ref{s:differential2} are no longer valid. 
These modified cumulants are defined in Sec.\ \ref{s:integrateda}
for integrated flow and in Sec.\ \ref{s:differentiala} for 
differential flow. 
As we shall see, the analytical expression of higher order 
cumulants become very lengthy, so that it is more convenient to 
work directly at the level of generating functions. 
As an illustration of our method, results of a simple 
Monte-Carlo simulation are given in Sec.\ \ref{s:montecarlo}. 

\subsection{Integrated flow}
\label{s:integrateda}

The key idea is that anisotropies in the detector 
acceptance can be handled much in the same way as anisotropies 
of the emitting source. The only difference is that the 
relevant coordinate system is the laboratory system in the 
first case, and the system associated with the 
reaction plane in the second case. 

Let us be more specific: until now, we have been working 
in the coordinate system associated with the reaction plane, 
i.e.\ with the emitting source.  
In this system, we used a cluster expansion to define 
direct $k$-particle correlations, of order $N^{1-k}$ relative 
to the uncorrelated $k$-particle distribution 
(Sec.\ \ref{s:multiparticle}). This cluster expansion 
allowed us to construct the ``connected moments'' of the 
distribution of $Q$, which were noted as $\connex{Q^kQ^{*l}}$ 
(Appendix \ref{s:b1}), of order $M^{1-k-l}$ relative 
to the corresponding moment $\mean{Q^kQ^{*l}}$. 
This decomposition was performed for an arbitrary source, 
but with an ideal detector. 

Exchanging the roles played by the source and the 
detector, the same reasoning applies if we work with 
an isotropic source and an imperfect detector, provided 
we use the coordinate system associated with 
the detector. We thus define the connected moments exactly 
in the same way, replacing $\phi$ by the measured $\bar\phi$ 
(see Sec.\ \ref{s:fourier}). Similarly, the flow vector $Q$ 
will be denoted $\bar Q$ when azimuthal angles are measured 
in the laboratory system, i.e.\ when $\phi_j$ is replaced
by $\bar\phi_j$ in the definition (\ref{qn}). 

I
f the acceptance is not perfect, 
averages such as $\langle e^{in\bar\phi} \rangle$ or
$\langle e^{in(\bar\phi_1+\bar\phi_2-\bar\phi_3)} \rangle$
no longer vanish. Thus, nondiagonal moments 
$\mean{\bar Q^k\bar Q^{*l}}$ 
with $k\not= l$ are also nonvanishing: 
there is no more cancellation due to isotropy, 
and all terms must be kept in the cumulant expansion. 
At order 2, for instance, the cumulants are defined as 
\begin{eqnarray}
\label{cumul2}
\cumul{\bar Q^2} & \equiv & \mean{\bar Q^2} - \mean{\bar Q}^2,\cr
\cumul{|\bar Q|^2} & \equiv & \mean{|\bar Q|^2}-\mean{\bar Q}\mean{\bar Q^*}.
\end{eqnarray}
These cumulants are of the same magnitude as when the acceptance 
is perfect, i.e.\ of order unity, while the moments $\mean{\bar Q^2}$ and
$\mean{|\bar Q|^2}$ scale like the multiplicity $M$ if the detector
is very bad. 
Note that at this order ($k+l=2$), taking the cumulant is 
equivalent to shifting the distribution of $\bar Q$ by its average value 
$\mean{\bar Q}$, as proposed in \cite{poskanzer98}. 

Higher order cumulants can be obtained in similar way as for 
an ideal detector. The only difference 
is that the simplifications due to isotropy no longer exist. 
Thus one cannot use expression (\ref{genfunc2}) for the generating function 
of the moments; one must use instead the more general expression 
(\ref{genfunc1}). The cumulant $\cumul{\bar Q^k \bar Q^{* l}}$ is 
therefore defined by 
\begin{equation} 
\sum_{k,l} {z^{*k} z^{l}\over k!\, l!} \cumul{\bar Q^k \bar Q^{* l}} =
\ln{\cal G}_0(z)=
\ln\mean{e^{z^*\bar Q + z\bar Q^*}}.
\label{logGa}
\end{equation}
Expanding the right-hand side to order $z^{*k} z^{l}$, one obtains 
the cumulant $\cumul{\bar Q^k \bar Q^{* l}}$ as a function of the 
measured moments $\langle\bar Q^{k'} \bar Q^{* l'}\rangle$ with 
$k'\le k$ and $l'\le l$. 
While the moment $\langle\bar Q^k \bar Q^{* l}\rangle$  is of 
magnitude $M^{(k+l)/2}$ for 
a bad detector, the corresponding cumulant  $\cumul{\bar Q^k \bar Q^{* l}}$ 
is of order $M^{(k+l)/2}N^{1-k-l}\sim M^{1-(k+l)/2}$. 

If the acceptance is not too bad, we assume that relation (\ref{geneq})
between the cumulants and the integrated flow is approximately 
preserved. 
The integrated flow can then obtained from the cumulants to 
order 2, 4, 6 using Eqs.\ (\ref{q2order}), (\ref{q4}) and 
(\ref{q6}), which we write again in the form: 
\begin{mathletters}
\label{intflow246}
\begin{eqnarray}
\mean{Q}^ 2&=& \cumul{|\bar Q|^2|}-1+O(1)
\pm \sqrt{1+2\langle Q\rangle^2\over N_{\rm evts}}
,\label{intflow246a}\\
\mean{Q}^ 4&=& -\cumul{|\bar Q|^4}-{1\over M}+O\left({1\over M}\right)
\pm 2\sqrt{1+4\langle Q\rangle^2+\langle Q\rangle^4+2\langle Q\rangle^6
\over N_{\rm evts}}
,\label{intflow246b}\\
\mean{Q}^ 6&=& {1\over 4}\cumul{|\bar Q|^6}-{1\over M^2}+
O\left({1\over M^2}\right)\pm {3\over 2\sqrt{N_{\rm evts}}},
\label{intflow246c}
\end{eqnarray}
\end{mathletters}
where, in the right-hand side of each equation, the last three 
terms stand for autocorrelations, 
systematic errors due to direct $2k$-particle correlations, 
and statistical errors due to the finite 
number of events (see Sec.\ \ref{s:statistical}), respectively.
Note that $\mean{Q}$ denotes the average value of $Q$ in the 
coordinate system associated with the reaction plane, i.e.\ what 
we call the ``integrated flow''. It must not be mistaken for 
$\mean{\bar Q}$ [see for instance Eq.\ (\ref{cumul2})], which denotes 
the average value in the laboratory coordinate system, and 
vanishes if the acceptance is perfect.   
Note also that only the ``diagonal cumulants'' 
$\cumul{|\bar Q|^{2k}}$ (i.e.\ with $k=l$) are related to the flow. 
These diagonal cumulants could equivalently be written 
$\cumul{|Q|^{2k}}$ since $Q$ and $\bar Q$ differ only by a phase. 
Other cumulants, $\cumul{Q^{k}Q^{*l}}$ with $k\not= l$, are not 
influenced by the flow and vanish except for statistical 
and systematic errors. 
They can therefore be used to estimate the magnitude of errors. 

The modified definition of higher order cumulants 
involves a large number of terms when the detector acceptance  is nonisotropic.
For instance, the fourth-order cumulant is obtained by expanding 
Eq.\ (\ref{logGa}) to order $z^2z^{*2}$: 
\begin{eqnarray}
\cumul{|\bar Q|^4}=
\mean{|\bar Q|^4} & - & 2\mean{\bar Q}\mean{\bar Q \bar Q^{* 2}} - 
2\mean{\bar Q^*}\mean{\bar Q^*\bar Q^2} - 2\mean{|\bar Q|^2}^2 - 
\mean{\bar Q^2}\mean{\bar Q^{* 2}}\cr
 & + & 8\mean{\bar Q}\mean{\bar Q^*}\mean{|\bar Q|^2} + 
2\mean{\bar Q}^2\mean{\bar Q^{* 2}} + 2\mean{\bar Q^*}^2\mean{\bar Q^2} - 
6\mean{\bar Q}^2\mean{\bar Q^*}^2 .
\label{q4a}
\end{eqnarray}
This equation replaces Eq.\ (\ref{eq2p-2p_c}) for an imperfect detector.
It shows that implementing acceptance corrections 
order by order can be very tedious since it involves a large number of terms. 

It is simpler to work 
directly with generating functions. 
Although this might seem to be more complicated, it is not unnatural since 
the generating functions constructed from experimental data have the 
same geometrical properties as the data, in particular regarding the detector 
acceptance. For instance, when the detector is isotropic, so is the 
generating function Eq.\ (\ref{genfunc2}). 

One can compute numerically 
the generating function of the cumulants ${\cal G}_0(x,y)$ 
at various points in the complex plane, 
then extract numerically the coefficients at a given order
by means of an interpolating polynomial. 
Let us be more specific: separating the real and imaginary 
part of the flow vector, we write it as 
\begin{eqnarray}
\bar Q_x &\equiv& \frac{1}{\sqrt{\sum_{j=1}^M w^2_j}} \sum_{j=1}^M w_j 
\cos(n\bar\phi_j),\cr
\bar Q_y &\equiv& \frac{1}{\sqrt{\sum_{j=1}^M w^2_j}} \sum_{j=1}^M w_j 
\sin(n\bar\phi_j).
\label{qnxy}
\end{eqnarray}
The generating function of the cumulants, defined by 
Eq.\ (\ref{logGa}), is a real-valued function: 
\begin{equation}
\label{gxy}
\ln {\cal G}_0(x,y)\equiv\ln\mean{e^{2x \bar Q_x+2y\bar Q_y}},
\end{equation}
where we have set $z=x+iy$. 

According to Eq.\ (\ref{logGa}), 
the cumulant to order $2k$, $\cumul{|\bar Q|^{2k}}$ is the 
coefficient of $(z z^*)^k=(x^2+y^2)^k$ in the power series 
expansion of this generating function, up to a factor $1 / (k!)^2$: 
\begin{equation}
\label{cumuli}
\ln{\cal G}_0(x,y)\equiv\sum_{k=1}^{\infty}
{\cumul{|\bar Q|^{2k}}\over (k!)^2}(x^2+y^2)^k,
\end{equation}
where we have kept only the relevant terms in the expansion. 
The cumulant can be obtained from the tabulated values of 
$\ln{\cal G}_0(x,y)$ using the interpolation formulas given 
in Appendix \ref{s:interpoli}. 

\subsection{Differential flow}
\label{s:differentiala}

When measuring differential flow, acceptance corrections can be implemented 
in the same way as for integrated flow. 
Flow is extracted using the same formulas as when the detector is perfectly 
isotropic in azimuth (Sec.\ \ref{s:differential}), 
without the simplifications allowed by isotropy. 
Therefore, one must take as the generating function of the cumulants 
${\cal C}_m(z)$ the general expression (\ref{genfuncdiff1}) instead of 
Eq.\ (\ref{genfuncdiff2}). 
We thus define the cumulants by 
\begin{equation}
\sum_{k,l} {z^{*k}z^l\over k!\, l!}
\cumul{\bar Q^k \bar Q^{* l} e^{im\bar\psi}}=
{\cal C}_m(z)\equiv 
\frac{\mean{e^{z^*\bar Q + z \bar Q^* + im\bar\psi} }}
{\mean{e^{z^*\bar Q + z \bar Q^*}}}
\label{geneqdiffa}
\end{equation}
where $\bar\psi$ denotes the azimuthal angle of the proton, 
measured in the laboratory coordinate system. 
This equation replaces Eq.\ (\ref{gendiffcumul}) 
for an imperfect detector. 
Expanding Eq.\ (\ref{geneqdiffa}) to order $z$ for $m=1$, we 
obtain for instance 
\begin{equation}
\label{q2diffa}
\cumul{\bar Q^* e^{i\bar\psi}}\equiv 
\mean{\bar Q^* e^{i\bar\psi}}-
\mean{\bar Q^*}\mean{e^{i\bar\psi}}
\end{equation}

We assume that the relation (\ref{geneqdiff}) 
between the cumulants and the differential flow, $v'_m$, 
is approximately preserved if the acceptance is not too bad. 
For $k=0$ and $k=1$, flow is then related to the 
cumulants by Eqs.\ (\ref{q2diff}) and (\ref{q4diff}) for $m=1$ 
and by Eqs.\ (\ref{v21bis}) and (\ref{q22diffflow}) for $m=2$. 
We rewrite  these formulas: 
\begin{mathletters}
\label{impl_diff11}
\begin{eqnarray}
\mean{Q} v'_1&=&\cumul{Q^{*}e^{i\psi}}
+O\left({1\over M^{1/2}}\right)\pm 
{1\over\sqrt{N'_{\rm evts}}},\label{impl_diff11a} \\
\mean{Q}^3 v'_1&=&-\cumul{|Q|^{2}Q^{*}e^{i\psi}}
+O\left({1\over M^{3/2}}\right)\pm 
{1\over\sqrt{N'_{\rm evts}}},\label{impl_diff11b}\\
\mean{Q}^2 v'_2&=&\cumul{Q^{*2}e^{2i\psi}}+
O\left({1\over M}\right)\pm 
{1\over\sqrt{N'_{\rm evts}}},\\
\mean{Q}^4 v'_2&=&-{1\over 2}\cumul{|Q|^{2}Q^{*2}e^{2i\psi}}+
O\left({1\over M^2}\right)\pm 
{1\over\sqrt{N'_{\rm evts}}},
\end{eqnarray}
\end{mathletters}
where, in the right-hand side of each equation, the second term 
represents the systematic error due to direct particle correlations,
while the last term is the statistical error due to the finite 
number of events. 
Note that only the cumulants $\cumul{Q^{k}Q^{*\,l}e^{im\psi}}$ 
with $l=k+m$ are related to the flow by Eqs.\ (\ref{impl_diff11}). 

We wish to recall here that the differential flow $v'_2$ might also have 
been obtained from the correlation between the azimuth of the proton and 
the event flow vector $Q_2$. 
As stated in Sec.\ \ref{s:differential2}, the only modification is a 
multiplication of all angles by 2, so that this does not change Eqs.\ 
(\ref{geneqdiffa}) and (\ref{cxy}). 
Therefore, $v'_2$ may be deduced from Eqs.\ (\ref{impl_diff11a}) 
and (\ref{impl_diff11b}) by the 
simple substitution of $v'_1$ and $\langle Q \rangle$ by $v'_2$ and 
$\langle Q_2 \rangle$ respectively.

As in the case of integrated flow, the modified 
definitions of the cumulants quickly involve a large number 
of terms when going to higher orders. Therefore, it is simpler 
in practice to extract the cumulants numerically from the 
generating function. 
For this purpose, 
one must tabulate numerically the real and imaginary parts of 
${\cal C}_m(z)$: 
\begin{eqnarray}
\label{cxy}
{\Re}[{\cal C}_{m}(x,y)]&=&
\frac{\mean{e^{2x \bar Q_x+2y\bar Q_y}\cos(m\bar\psi)}}
{\mean{e^{2x \bar Q_x+2y\bar Q_y}}},\cr
{\Im}[{\cal C}_{m}(x,y)]&=&
\frac{\mean{e^{2x \bar Q_x+2y\bar Q_y}\sin(m\bar\psi)}}
{\mean{e^{2x \bar Q_x+2y\bar Q_y}}}.
\end{eqnarray}
Keeping only the terms with $l=k+m$ which are related to the 
flow, the generating function (\ref{geneqdiffa}) becomes   
\begin{equation}
\label{devcm}
{\cal C}_m(z)=\sum_{k=0}^{\infty} {\cumul{|Q|^{2k}Q^{*m}e^{im\psi}}
\over k! (k+m)!} z^{*k} z^{k+m}.
\end{equation}
Interpolation methods to calculate the cumulants 
$\cumul{|Q|^{2k}Q^{*m}e^{im\psi}}$
as a function of the tabulated 
values of the generating function, are explained in detail in 
Appendix \ref{s:interpold}. 

\subsection{Results of a Monte-Carlo simulation}
\label{s:montecarlo}

We have tested our method with a simple Monte-Carlo simulation.
Particles have been generated randomly with the distribution 
\begin{equation}
{{\rm d}N\over {\rm d}\phi}\propto 1+2v_1\cos\phi+2v_2\cos(2\phi). 
\end{equation}
The value of the integrated directed flow, which we tried 
to reconstruct, was fixed to $v_1=0.03$, corresponding roughly (up to a sign) 
to the value measured at SPS for pions \cite{na49}. 
We have taken various values of $v_2$, in order to probe the interference 
between both harmonics, discussed in Sec.\ \ref{s:multiparticle2}. 

In a first step, we do not simulate nonflow correlations 
between the particles.
In order to take into account the effect of detector 
inefficiencies, we have 
assumed that all particles are detected, except in a blind 
azimuthal sector of size $\alpha$. 
The simulation has been performed with $N_{\rm evts}=200000$ events, 
and a multiplicity $M=200$ for each event. For simplicity, we have 
assumed that exactly 200 particles are detected in each event. 
Fluctuations in $M$ should not influence the results, as explained 
in Sec.\ \ref{s:sub3a3}. 
With these values, the optimal sensitivity for the integrated
flow is obtained for 
$k=2$ according to Eq.\ (\ref{err-stat2}), i.e.\ by taking 
the fourth order cumulant. 
We therefore reconstruct the flow using Eq.\ (\ref{intflow246b}).

With the values we have chosen,
$\mean{Q}=v_1\sqrt{M}\simeq 0.42<1$, so that traditional 
methods might fail, as stated before. 
Within our method, 
the statistical error on $v_1$, calculated with 
Eq.\ (\ref{erstat4}), is of the order of $0.14\%$. 
Since direct correlations between particles are not simulated, 
the only systematic error comes from detector inefficiencies 
and the higher harmonic $v_2$. 

Results are shown in the table below. 
The table gives the reconstructed $v_1$ as a function of the size of the
blind angle $\alpha$, and the higher harmonic $v_2$.

\begin{table}[htbp]
\begin{center}
\begin{tabular}{|l|c|c|c|c|c|}
\hline
\cline{2-6}
\multicolumn{1}{l|}{}
& $\alpha= 0^{\circ}$ & $\alpha=45^{\circ}$ & $\alpha=$90$^{\circ}$ & 
$\alpha=$135$^{\circ}$ & $\alpha=$180$^{\circ}$\\
\hline
$v_2=0$ &3.04 & 3.10 & 3.11& 2.91 & 2.11\\
\hline
$v_2=3$ &2.83 & 2.85 & 2.98 & 2.78 & 2.57\\
\hline
$v_2=6$ &2.65 & 2.82& 2.78 & 3.55 & 4.24\\
\hline
$v_2=-3$ &3.30 & 3.22& 3.23 & 2.99 & 2.57 \\
\hline
\end{tabular}
\end{center}
\caption{Results of a Monte-Carlo simulation. 
All  values of $v_1$ and $v_2$ are given in $\%$. }
\end{table}
If $v_2=0$, the reconstructed value is compatible with the theoretical
value within statistical errors, except for the highest value of $\alpha$, 
i.e.\ when the detector covers only half of the range in azimuth. 
Therefore, errors due to acceptance imperfections are under good 
control.  

The systematic error from higher harmonics, on the other hand, 
is far from negligible. The limits of applicability of our 
method, given by Eq.\ (\ref{app-v2}), are here $-0.43<v_2<0.07$. 
We have checked these bounds numerically. 
The value $v_2=0.06$ is very close to the upper bound. 
However, the corresponding relative error on $v_1$ is only 12\%
with an ideal detector. 

In a second step, we simulate nonflow correlations: for 
simplicity, we do this assuming that particles are emitted in 
pairs, both particles in a pair having exactly the same azimuthal
angle. This would be the case for the two-body decay of a very 
fast resonance. Taking the same numerical values as above, 
the standard method, corresponding to Eq.\ (\ref{intflow246a}), 
gives $v_1=7.7\%$: it fails, as expected, overestimating 
the flow by more than a factor of 2.  
On the other hand, the fourth-order formula
(\ref{intflow246b}), which eliminates two-particle nonflow 
correlations, gives $v_1=3.1\%$, in much better agreement 
with the theoretical value.

\section{Summary}
\label{s:summary}

We have proposed in this paper a new method for the flow 
analysis, which is 
more sensitive than traditional methods to small anisotropies 
of the azimuthal distributions. In this section, we summarize the 
procedure which should be followed in practice.  

The first step consists in measuring the ``integrated flow,'' as explained 
in Sec.\ \ref{s:integrated}.  
This corresponds to the 
problem of the reaction plane determination in the standard 
flow analysis. 
One first constructs, event by event, the flow vector 
$\bar Q_n$ defined by Eq.\ (\ref{qnxy}), where the $\bar\phi_j$ 
are the azimuthal angles of the particles in the laboratory 
coordinate system. 
The weight $w_j$ is chosen as explained in Sec.\ \ref{s:sub3d2}; 
ideally, it should be taken equal to the differential flow
$v_n(p_T,y)$, 
i.e.\ proportional to $p_T^n$, and even (resp.\ odd) in the rapidity $y$ for 
even (resp.\ odd) $n$. 
Alternatively, one may choose the simpler version with unit weights (\ref{qn}). 
The value of $n$ depends on the system under study: 
up to energies of 10~GeV per nucleon, one usually works with 
$n=1$, i.e.\ with $Q_1$ \cite{danielewicz85,E877,E877bis}. 
At SPS, directed flow is so small 
that a better accuracy is obtained by working directly with 
the second harmonic, i.e.\ by constructing $Q_2$ \cite{na49}. 
Then, only even harmonics can be measured. 
Most of this paper has been written assuming $n=1$. 
In order to generalize the results to the case $n=2$, one need only 
multiply {\it all\/} azimuthal angles by 2. 

Measuring the integrated flow amounts to measuring the average 
value of the flow vector, $\mean{Q_n}$,
in the coordinate system where the reaction plane is fixed. 
The average value $\mean{Q_n}$ is of order 
$v_n\sqrt{M}$ (it is even equal to that value if one is working with 
unit weights), where $v_n$ is the Fourier harmonic of 
order $n$, and $M$ the number of particles used in the 
flow analysis. 
As explained in Sec.\ \ref{s:integrated}, the 
integrated flow $\mean{Q_n}$ is obtained from the 
cumulant $\cumul{|Q_n|^{2k}}$, which removes 
nonflow correlations up to order  $2k$, 
the standard method corresponding to the lowest order, 
$k=1$. 
The value of $k$ is chosen so as to obtain the 
best sensitivity. It results from a balance between 
systematic and statistical errors, and depends both 
on the number of events $N_{\rm evts}$ available
for the flow analysis, and on the number of particles used 
to determine the reaction plane in each event, $M$. 
The optimal order $k$ is then given by Eq.\ (\ref{err-stat2}). 
However, performing measurements with other values of $k$ 
does not cost much and provides a useful comparison. 

The cumulant $\cumul{|Q_n|^{2k}}$ is 
a combination of the moments of the distribution of $\bar Q_n$, 
i.e.\ it is expressed as a function of the measured moments 
$\mean{\bar Q_n^l \bar Q_n^{*m}}$,
with $l\le k$ and $m\le k$. 
In this paper, we have used the formalism of generating functions 
to derive the corresponding formulas at arbitrary order. 
As explained in Sec.\  \ref{s:acceptance}, this is not only 
an elegant formalism: it is also the simplest way to 
calculate the cumulants numerically from experimental data. 
For this purpose, one tabulates the generating function 
${\cal G}_0(x,y)$, defined by Eq.\ (\ref{gxy}), at various 
points in the $(x,y)$ plane. 
In this equation, the brackets denote an average over 
the whole sample of events. 
The cumulant $\cumul{|Q_n|^{2k}}$ is 
then obtained by extracting numerically the coefficient 
in front of $(x^2+y^2)^k$ in the power series expansion 
of $\ln {\cal G}_0(x,y)$, as explained in Sec.\ \ref{s:integrateda}. 
The integrated flow $\mean{Q_n}$ is finally obtained from 
the cumulant using Eqs.\ (\ref{intflow246}). 

The value of $\mean{Q_n}$ is the important parameter 
in the flow analysis, since it determines the accuracy of 
the reconstruction of azimuthal distributions. 
If $\mean{Q_n}> 1$, the flow 
can easily be studied with traditional methods, although 
the present method should give more accurate results. 
If $\mean{Q_n}<1$, on the other hand, standard 
methods fail, while our method still works.

The second step in the flow analysis is to perform 
detailed measurements of the flow 
coefficient $v'_m$ for a particle of given rapidity and 
transverse momentum, i.e.\ differential flow. 
The coefficient $v'_m$ can be obtained from the comparison of the 
azimuth of the particle under study with an event flow vector, 
which can be either $Q_m$, calculated with the 
same harmonic, or a $Q_n$, calculated with a 
different harmonic, provided $m$ is a 
multiple of $n$. For instance, $v'_2$ can be measured with respect 
to $Q_1$ or $Q_2$, as explained in \cite{poskanzer98}.
We show in Sec.\  \ref{s:differential2} that it is the value of 
$\mean{Q_n}$ which determines the accuracy on the measurement 
of $v'_m$. 
Therefore, $n$ should be
chosen so that $\mean{Q_n}$ 
be as large as possible. For instance, at RHIC
where $v_2$ is expected to be much larger than $v_1$, 
$v'_2$ should be measured with $Q_2$ rather than with $Q_1$, as is 
already the case at SPS \cite{na49}.
In the text, we have assumed $n=1$. If one uses $Q_2$, then $m$ 
must be replaced by $2m$ everywhere in our equations. 

As the integrated flow, the differential flow $v'_m$ is obtained 
from a cumulant $\cumul{|Q|^{2k} Q^{*m} e^{im\psi}}$ which  
eliminates nonflow correlations up to an arbitrary 
order $2k+m$, the standard analysis corresponding to the case $k=0$. 
Here again, the best choice of $k$ is the one which leads to the smallest 
error: its value is given by Eq.\ (\ref{kdiff}). 
In order to measure the cumulants, one first 
tabulates the generating function (\ref{cxy}) at various 
points in the complex plane. 
The cumulant is then obtained by extracting the coefficient 
proportional to $z^{*k} z^{k+m}$ in the power series 
expansion of the generating function, as explained in 
Sec.\ \ref{s:differentiala}. Finally, the differential flow $v'_m$ 
is related to the cumulants by Eqs.~(\ref{impl_diff11}).  

A limitation of our method at a given order is the possible 
interplay of higher harmonics in the measurement. 
For instance, Eq.\  (\ref{c4order}) shows that in the 
fourth-order cumulant, the second 
harmonic $v_{2n}$ interferes with $v_n$. 
More precisely, $|v_{2n}|$ must be small compared with $Mv_n^2$ 
(see Eq.\ (\ref{limitv2})). 
This limitation means that the method should be used with much 
care when extracting the directed flow ($n=1$) at RHIC and 
LHC~\cite{inprogress}, since it is expected to be much smaller than 
elliptic flow. 
On the other hand, in the case $n=2$, there should be no 
problem since $v_2$ is much larger than $v_4$. 

While higher harmonics or statistical errors may limit the use of the
method, there is no problem with the acceptance of detectors. As a
matter of fact, the required corrections appear in a natural way in 
the method, at all orders, from a modification of the generating
equation which is the same for all detectors. In particular, the sensitivity 
remains unchanged when acceptance corrections are taken into account, 
so that choosing the order in the expansion of the generating equation does 
not depend on that problem. 

Most of our results have been established in the limit 
where azimuthal anisotropies are weak. For this reason, our 
method seems to be more adapted to ultrarelativistic energies, i.e.\ 
at SPS energies and beyond, where $v_1$ and $v_2$ are usually 
less than $0.1$. 
In particular, it should be very useful in the forthcoming 
flow analyses at the Brookhaven Relativistic Heavy Ion Collider.

\acknowledgements

We thank Art Poskanzer for helpful comments on the first version 
of this paper, and Raimond Snellings and Sergei Voloshin for 
stimulating discussions. 
We also thank Aihong Tang for correcting some misprints.
\appendix

\section{Detailed study of the four-particle azimuthal correlation}
\label{s:app-q4}

In Sec.\ \ref{s:app-c4c}, we calculate the cumulant of the four-particle
azimuthal correlation, introduced in Sec.\ \ref{s:multiparticle2}.
Then, in Sec.\ \ref{s:app-diag}, we calculate the fourth order
cumulant of the $Q$ distribution, introduced in Sec.\ \ref{s:integrated2}.

\subsection{Cumulant of the four-particle correlation}
\label{s:app-c4c}

The cumulant of the four-particle azimuthal distribution 
has been defined by Eq.\ (\ref{defcumul4}) when the source is 
isotropic. We set $n=1$ for simplicity:
\begin{equation}
\label{c4cdef}
\cumul{e^{i(\phi_1+\phi_2-\phi_3-\phi_4)}} \equiv
\mean{e^{i(\phi_1+\phi_2-\phi_3-\phi_4)}} -
\mean{e^{i(\phi_1-\phi_3)}} \mean{e^{i(\phi_2-\phi_4)}} -
\mean{e^{i(\phi_1-\phi_4)}} \mean{e^{i(\phi_2-\phi_3)}}.
\end{equation}
Here, we want to evaluate the right-hand side of this 
equation when the source is no longer isotropic.

\begin{figure}[h]
\centerline{\includegraphics*[width=0.99\linewidth]{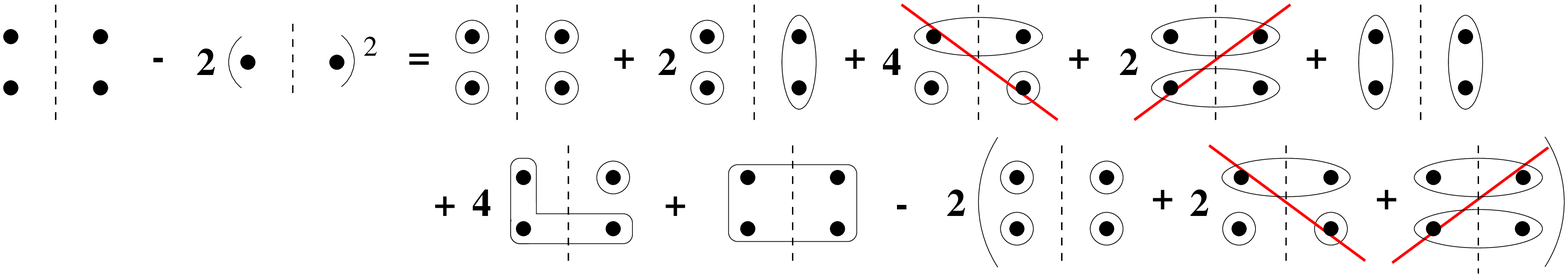}}
\caption{Expansion into connected parts of the cumulant of the four-particle 
azimuthal correlation. Dots on the left (resp.\ right) of the dashed line 
represent $e^{i\phi}$ (resp.\ $e^{-i\phi}$). 
}
\label{fig:fig2p-2p_c}
\end{figure}

In order to do so, we expand the four-particle distribution 
into connected parts, as explained in Sec.\ \ref{s:multiparticle}. 
Using the diagrammatic representation introduced there, 
the quantity in Eq.\ (\ref{c4cdef}) can be decomposed as 
in Fig.\ \ref{fig:fig2p-2p_c}. 
The diagrams in Fig.\ \ref{fig:fig2p-2p_c} stand for:
\begin{eqnarray}
\label{c4cdev}
\mean{e^{i(\phi_1+\phi_2-\phi_3-\phi_4)}} - 2\mean{e^{i(\phi_1-\phi_3)}}^2 = 
& - & v_1^4 + 2\,v_1^2 \connex{e^{-i(\phi_3+\phi_4)}} + 
\connex{e^{i(\phi_1+\phi_2)}} \connex{e^{-i(\phi_3+\phi_4)}} \cr 
& + & 4\,v_1\connex{e^{\pm i(\phi_1+\phi_2-\phi_3)}} +
\connex{e^{i(\phi_1+\phi_2-\phi_3-\phi_4)}} .
\end{eqnarray}
Note that the direct two-particle correlations $\connex{e^{i(\phi_1-\phi_3)}}$ 
are automatically removed.
In the isotropic case, only the connected part of the correlation, i.e.\ 
$\connex{e^{i(\phi_1+\phi_2-\phi_3-\phi_4)}}$,
remains in the right-hand side of Eq.\ (\ref{c4cdev}). 

Let us now enumerate the orders of magnitude of the different terms in
the right-hand side of Eq.\ (\ref{c4cdev}).
As stated above, all terms but the last vanish in the isotropic case: indeed, 
$\connex{e^{i(\phi_1+\phi_2-\phi_3)}}$ and $\connex{e^{\pm i(\phi_1+\phi_2)}}$
are not invariant under the transformation 
$\phi_j \rightarrow \phi_j+\alpha$, where $\alpha$ is any angle. Therefore, 
it seems reasonable to consider that these terms are proportional to $v_1$ or
$v_2$, depending on whether a factor $e^{\pm i\alpha}$ or
$e^{\pm 2i\alpha}$ appears under the previous transformation. 
Furthermore, since we consider here connected $k$-particle
correlations, they behave like $O(1/N^{k-1})$ [see Sec.\ \ref{s:multiparticle}]. 
More precisely,
\begin{equation}
\connex{e^{i(\phi_1+\phi_2-\phi_3)}} = O\left(\frac{v_1}{N^2}\right),
\qquad
\connex{e^{\pm i(\phi_1+\phi_2)}} = O\left(\frac{v_2}{N}\right).
\end{equation}
Note that the second term in the right-hand side of Eq.\ (\ref{c4cdev})
is smaller than either the first or the third terms.

Finally, the order of magnitude of the right-hand side of 
Eq.\ (\ref{c4cdef}) is $v_1^4+O(v_2^2/N^2+1/N^3)$.
We have neglected $v_1^2/N^2$ 
since it is smaller than either $v_1^4$ or $1/N^3$.

\subsection{Calculation of the cumulant $\cumul{|Q|^4}$}
\label{s:app-diag}

In this section, we derive the order of magnitude of the 
fourth order cumulant of the $|Q|^2$ distribution, 
defined by Eq.\ (\ref{eq2p-2p_c}).  
From the definition of the event flow vector Eq.\ (\ref{qn}), one
obtains 
\begin{equation}
\cumul{|Q|^4 }=  \frac{1}{M^2} 
\sum_{j,k,l,m} \left( \mean{e^{i(\phi_j+\phi_k-\phi_l-\phi_m)}} - 
\mean{e^{i(\phi_j-\phi_l)}} \mean{e^{i(\phi_k-\phi_m)}} - 
\mean{e^{i(\phi_j-\phi_m)}} \mean{e^{i(\phi_k-\phi_l)}} \right). 
\label{q4cdev}
\end{equation}
In the above sum, one may distinguish nondiagonal terms, 
when all four indices are different, and diagonal terms, for 
which at least two indices are equal. 

Nondiagonal terms correspond precisely to the cumulant of the 
four-particle correlation. 
The corresponding contribution, evaluated in 
Sec.\ \ref{s:app-c4c}, must be multiplied by the combinatorial 
factor $M(M-1)(M-2)(M-3)\sim M^4$. 
With the factor $1/M^2$ in front of the sum in Eq.\ (\ref{q4cdev}), 
the contribution of nondiagonal terms to $\cumul{|Q|^4 }$ is 
of order $M^2 v_1^4+O(v_2^2+1/N)$. 

We are now going to show that diagonal terms give a contribution 
at most of the same order as nondiagonal terms. 
Let us enumerate the various diagonal terms:

\begin{enumerate}[i)]
\item If $j=k=l=m$, each term in the sum is equal to $-1$. 
This is the contribution which we call ``autocorrelations''. 
Multiplying by a combinatorial factor $M$ and by 
the factor $1/M^2$ in Eq.\ (\ref{q4cdev}), the corresponding 
contribution is exactly $-1/M$.

\item When three indices are identical while the fourth is different, 
i.e.\ in $4M(M-1)$ cases, the difference in Eq.\ (\ref{q4cdev}) reduces 
to $-\mean{e^{i(\phi_1-\phi_2)}}$. Using Eq.\ (\ref{c2order}),
this contribution is of order $-4v_1^2+O(1/N)$. Although this
contribution is a two-particle correlation, it is suppressed by the
combinatorial factor: $v_1^2$ is much smaller than the term $M^2v_1^4$ 
which appears in the cumulant of the four-particle azimuthal
correlation (see Sec.\ \ref{s:app-c4c}). Therefore, this
contribution will be negligible. 

\item Let us consider the cases when the indices are equal two by two.
\begin{itemize}
\item If $j=k$ and $l=m$ but $j\neq l$, which occurs $M(M-1)$ times,
the difference is given by 
\begin{equation}
\mean{e^{2i(\phi_1-\phi_3)}} - 2\mean{e^{i(\phi_1-\phi_3)}}^2 =  
v_2^2 + \connex{e^{2i(\phi_1-\phi_3)}} -
2\left(v_1^2 + \connex{e^{i(\phi_1-\phi_3)}}\right)^2. 
\end{equation}
The order of magnitude is then $v_2^2 + O(1/N)$. Here, we have 
neglected terms of order $v_1^2/N$ and $1/N^2$, smaller than $1/N$;
the term $v_1^4$ is smaller by a combinatorial 
factor $1/M^2$ than the similar contribution of nondiagonal terms. 
Note that the higher harmonic $v_2$ contributes here. We
shall see below that these higher harmonics can limit the use 
of our method. 

\item The $2M(M-1)$ cases \{$j=m$ and $k=l$ but $j \neq l$\} or 
\{$j=l$ and $k=m$ but $k \neq l$\} yield a contribution  
$-\mean{e^{i(\phi_1-\phi_3)} }^2$. Its order of magnitude is 
$-2v_1^4+O(1/N^2)$, negligible compared to nondiagonal terms. 

\end{itemize}

\item There are two cases when three indices are different:
\begin{itemize}
\item If $j=l$ or $j=m$ or $k=l$ or $k=m$, while the two remaining
indices are different, the contribution is $-\mean{e^{i(\phi_1-\phi_3)} } ^2$, to be multiplied by a combinatorial
factor $4M(M-1)(M-2)$. Thus, the order of magnitude is
$M[-4v_1^4+O(1/N^2)]$ and this contribution is suppressed by a factor
$1/M$ with respect to the cumulant of the four-particle correlation.

\item If the two identical indices are either $(j,k)$ or $(l,m)$, the
combinatorial factor is $2M(M-1)(M-2)$, which multiplies a term
$\mean{e^{\pm i(2\phi_1-\phi_3-\phi_4)} } -2\mean{e^{i(\phi_1-\phi_3)} } ^2$. 
Using Eq.\ (\ref{fc3}), the
three-particle correlation $\mean{e^{i(2\phi_1-\phi_3-\phi_4)}
}$ can be expanded as
\begin{eqnarray}
\label{app-c3}
\mean{e^{i(2\phi_1-\phi_3-\phi_4)}} & = & 
\mean{e^{2i\phi_1}} \mean{e^{-i\phi_3}} \mean{e^{-i\phi_4}} +
\connex{e^{i(2\phi_1-\phi_3)}} \mean{e^{-i\phi_4}} + 
\connex{e^{i(2\phi_1-\phi_4)}} \mean{e^{-i\phi_3}}
\cr
& & +\mean{e^{2i\phi_1}} \connex{e^{-i(\phi_3+\phi_4)}} + 
\connex{e^{i(2\phi_1-\phi_3-\phi_4)}} \cr
& = & v_2 v_1^2 + O\left(\frac{v_1^2}{N}\right) +
O\left(\frac{v_2^2}{N}\right) + O\left(\frac{1}{N^3}\right).
\end{eqnarray}
The second term in the difference, 
$-2\mean{e^{i(\phi_1-\phi_3)}}^2$, gives a contribution of 
$-2v_1^4+O(1/N^2)$.
Finally, since terms such as $v_1^4$, $v_1^2/N$ are suppressed because 
of the combinatorial factor, the contribution in this case is 
$2Mv_2 v_1^2+O(Mv_2^2/N)+O(M/N^2)$.
\end{itemize} 
\end{enumerate}
We shall assume that the total multiplicity in the collision $N$ and 
the number $M$ of particles used to calculated the flow vector are
large and of the same order of magnitude. Then, we find that the 
contribution of the diagonal terms is 
$-1/M+v_2^2+2Mv_1^2 v_2 + O(v_2^2+1/M)$.

All in all, when we add the contributions of diagonal 
and nondiagonal terms, we obtain the following result: 
\begin{equation}
\label{app-q4order}
\cumul{|Q|^4} = 
-\frac{1}{M} - M^2 v_1^4 + 2Mv_1^2 v_2 + v_2^2 + 
O(v_2^2)+O\left(\frac{1}{M}\right).
\end{equation}
The first term in the right-hand side coresponds to 
autocorrelations, the last two terms are due to nonflow 
correlations, and the three remaining terms arise from flow. 
One would like $-M^2v_1^4$ to be the dominant flow term. 
However, higher harmonics, i.e.\ $v_2$, also contribute. 
If $v_2$ is large enough, it  may even reverse the sign of the 
contribution of flow to $\cumul{|Q|^4}$. 
This does not happen provided $v_2$ lies in the following 
interval~:
\begin{equation}
\label{app-v2}
-Mv_1^2(\sqrt{2}+1) < v_2 < Mv_1^2(\sqrt{2}-1).
\end{equation}
We have checked these bounds with our 
Monte-Carlo simulation, see Sec.\ \ref{s:montecarlo}.

\section{A generating equation for the integrated flow}
\label{s:app-integrated}

In this Appendix, we first construct the cumulants of the 
distribution of $|Q|$, which we note $\cumul{|Q|^{2k}}$, 
as a function of the measured moments 
$\mean{|Q|^{2k}}$ (Secs.\ \ref{s:b1} and \ref{s:b2}). 
Then, we relate the cumulants to the integrated
flow $\mean{Q}$ (Sec.\ \ref{s:b3}), and show how to 
remove autocorrelations at all orders (Sec.\ \ref{s:b4}).

\subsection{Cluster decomposition of the moments}
\label{s:b1}

We have shown in Sec.\ \ref{s:multiparticle} how the $k$-particle
momentum distribution can be decomposed, in a coordinate frame where the 
reaction plane is fixed, into a sum of terms involving lower order 
distributions ($k'$ particles with $k'<k$), plus a ``connected'' term 
of relative order $1/N^{k-1}$. 
This decomposition also applies to the moments of the distribution of 
the event flow vector $Q$ defined by Eq.\ (\ref{qn}). 
As pointed out in Sec.\ \ref{s:integrated2}, moments 
of order $k$ involve $k$-particle azimuthal correlations. 
This allows us to write a series of equations similar to 
Eqs.\ (\ref{fc2}) and (\ref{fc3}):
\begin{mathletters}
\label{qcumulants}
\begin{eqnarray}
\mean{Q} & = & \connex{Q}\label{qcumulantsa}\\
\mean{Q^2} & = & \connex{Q}^2 + \connex{Q^2}\label{qcumulantsb}\\
\mean{QQ^*} & = & \connex{Q} \connex{Q^*} + \connex{QQ^*}\label{qcumulantsc}\\
\mean{Q^3} & = & \connex{Q}^3 + 3\connex{Q}\connex{Q^2} + \connex{Q^3}
\label{qcumulantsd}\\
\mean{Q^2Q^*} & = & \connex{Q}^2\connex{Q^*} + 2\connex{Q}\connex{QQ^*} +
\connex{Q^2}\connex{Q^*}+ \connex{Q^2Q^*}, {\rm\ etc.} \label{qcumulantse}
\end{eqnarray}
\end{mathletters}
In these equations, the subscript $c$ denotes ``connected'' moments.
The connected moment of order $k$ is of magnitude $M^{1-k/2}$: a factor 
$M^{1-k}$ comes from the fact that it involves direct $k$-particle correlations 
(see Sec.\ \ref{s:multiparticle}), and a factor $M^{k/2}$ from 
the scaling of $Q$ with the number of particles like
$\sqrt{M}$, see Eq.\ (\ref{qn}). 

\begin{center}
\begin{figure}[ht!]
\centerline{\includegraphics*[width=0.63\linewidth]{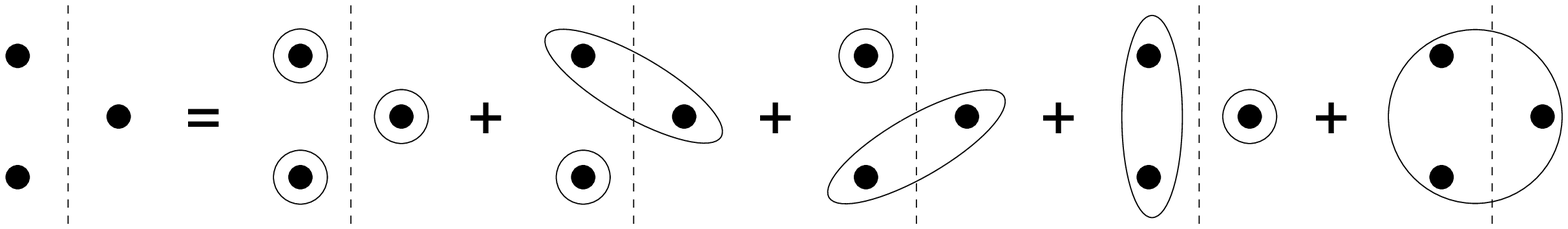}}
\medskip
\caption{Decomposition of $\mean{Q^2 Q^*}$ in connected parts,
see Eq.\ (\ref{qcumulantse}). Dots on the left of the dashed
line represent factors of $Q$ while dots on the right represent 
factors of $Q^*$. Circled subsets correspond to connected moments.}
\label{fig:fig2p-1p}
\end{figure}
\end{center}

The expansion of a given moment $\mean{Q^k Q^{* l}}$ 
in connected parts can be represented graphically by the 
expansion of a $(k+l)$-point diagram into connected diagrams. 
This is similar to the decomposition of  
the $k$-particle distribution in Figs.\ \ref{fig:fig2p} and 
\ref{fig:fig3p}. 
To be more specific, the decomposition of $\mean{Q^k Q^{* l}}$
is represented by drawing $k$ dots of one type corresponding to powers of $Q$
and $l$ dots of another type corresponding to powers of $Q^*$. 
One then takes all possible partitions of this set of $k+l$ points.
To each subset of points one associates the corresponding 
connected moment. 
The contribution of a given partition is
the product of the contributions of each subset. 
Finally, $\mean{Q^k Q^{* l}}$ is the 
sum of the contributions of all partitions. 
Figure \ref{fig:fig2p-1p} represents, as an example, the decomposition 
of $\mean{Q^2 Q^*}$. 

The connected moments can be expressed as a function of the moments by 
inverting Eqs.\ (\ref{qcumulants}) order by order. However, 
this procedure is very tedious. 
An elegant and compact way to express moments of arbitrary order in 
terms of the connected parts, and to invert these relations, consists 
in using generating functions. 
The generating function of the moments is a function of the complex 
variable $z$ which is defined as 
\begin{equation}
{\cal G}_0(z) = \mean{e^{z^*Q + zQ^*} }
=\sum_{k,l} {z^{*k} z^{ l}\over k!\, l!} \mean{Q^k Q^{* l}},  
\label{genfunc1}
\end{equation}
where $k$ and $l$ go from $0$ to $+\infty$, and 
the brackets denote an average over many events. 
It is well known in graph theory that the generating function 
of connected diagrams is the logarithm of the generating function 
of all diagrams~\cite{domb-green}. 
Therefore, the generating function of the connected moments is the logarithm 
of the generating function of the moments~\cite{vanKampen}:
\begin{equation}
\sum_{k,l} {z^{*k} z^{ l}\over k!\, l!} \connex{Q^k Q^{* l}}
=\ln{\cal G}_0(z). 
\label{logG}
\end{equation}
The normalization coefficient $1/k!\,l!$ has been chosen such that 
$\connex{Q^k Q^{* l}}$ appears with a unit coefficient 
in the expansion of $\mean{Q^k Q^{* l}}$, as in Eq.\ (\ref{qcumulants}).  
Expanding Eqs.\ (\ref{genfunc1}) and (\ref{logG}) to order $z^{*2}z$, 
one finds for instance 
\begin{eqnarray}
\label{qicumulants}
\connex{Q^2Q^*} & = & \mean{Q^2Q^*} - \mean{Q^2}\mean{Q^*}
- 2\mean{Q}\mean{QQ^*} + 2\mean{Q}^2\mean{Q^*},
\end{eqnarray}
which can be checked by inverting Eqs. (\ref{qcumulants}) order by
order. 
Note that we are working in a coordinate system where the reaction
plane correponds to the $x$-axis, and is unknown. In this coordinate system, 
the generating function (\ref{genfunc1}) is not a measurable quantity.

\subsection{Isotropic source}
\label{s:b2}

We now consider specifically an isotropic source, i.e.\ without flow. 
In that case, the moment $\mean{Q^k Q^{* l}}$ vanishes if $k\not=l$. 
The connected parts $\connex{Q^k Q^{* l}}$ enjoy the same property. 
Therefore, in the diagrammatic expansion, one only retains 
terms containing as many powers of $Q$ as of $Q^*$, i.e.\ 
as many dots on the left as on the right. 
The quantity represented in Fig.\ \ref{fig:fig2p-1p} does not 
satisfy this property, and therefore it vanishes. 
A decomposition with nonvanishing terms is represented in 
Fig.\ \ref{fig:fig2p-2p}. 

Keeping only the terms $k=l$, the generating function
 (\ref{genfunc1}) becomes 
\begin{equation}
{\cal G}_0(z) = \sum_{k=0}^\infty \frac{|z|^{2k}}{(k!)^2} 
\mean{|Q|^{2k}} = \mean{I_0(2|zQ|) },
\label{genfunc2}
\end{equation}
where $I_0$ is the modified Bessel function of order $0$. 
Note that now the generating function ${\cal G}_0$ itself is isotropic, since 
${\cal G}_0(z)={\cal G}_0(ze^{i\alpha})$. The consequence is 
that it can be evaluated in the laboratory coordinate system
rather than in the coordinate system associated with the 
reaction plane: it thus becomes a measurable quantity. 
We define the cumulants through 
\begin{equation}
\sum_{k=1}^\infty \frac{|z|^{2k}}{(k!)^2} \cumul{|Q|^{2k} } \equiv 
\ln {\cal G}_0(z)  = \ln \mean{I_0(2|zQ|) }. 
\label{genfunc3}
\end{equation}
They coincide with the connected moments $\connex{|Q|^{2k}}$ 
defined in Eq.\ (\ref{logG}) 
if the source is isotropic.
Note that for an isotropic system, the raw moment $\mean{|Q|^{2k}}$ 
is of order unity, as noted in Sec.\ 
\ref{s:integrated2}. The corresponding cumulant 
$\cumul{|Q|^{2k}}$ is of order $M^{1-k}$. 
Eq.\ (\ref{genfunc3}) corresponds to Eq.(\ref{geneqi}), where 
we have set $x=|z|$. 

\subsection{Flow}
\label{s:b3}

Let us now calculate the cumulants in the case of collisions with flow. 
Neglecting for simplicity nonflow correlations between particles, we can 
write $\mean{Q^k Q^{* l}}=\mean{Q} ^k
\mean{Q^{*}}^l =\mean{Q}^{k+l}$. The generating 
function (\ref{genfunc1}) thus becomes  
\begin{equation}
{\cal G}_0(z) = e^{(z+z^*) \mean{Q }}.
\label{denom}
\end{equation}
Now, we want to compare with the experimental value of ${\cal G}_0(z)$, which 
is measured in the laboratory coordinate system where the azimuth 
of the reaction plane $\phi_R\neq 0$. 
The generating function in this coordinate system is deduced 
from Eq.\ (\ref{denom}) by the substitution $z \to ze^{i\phi_R}$. 
Averaging the new expression over all possible $\phi_R$, under the 
assumption that the distribution of $\phi_R$ is uniform, one obtains: 
\begin{equation}
{\cal G}_0(z) = \frac{1}{2\pi} 
\displaystyle \int_0^{2\pi} \! e^{(ze^{i\phi_R} + z^*e^{-i\phi_R}) \mean{
Q }} \, {\rm d}\phi_R 
=I_0(2|z|\mean{Q }). 
\label{genfuncflow}
\end{equation}
Gathering the results obtained in Eqs.\ (\ref{genfunc3}) 
and (\ref{genfuncflow}), we obtain:
\begin{equation}
\sum_{k=1}^\infty \frac{|z|^{2k}}{(k!)^2} \cumul{|Q|^{2k} } = 
\ln{\cal G}_0(z)=
\ln I_0(2|z|\mean{Q }).
\label{flow+corr}
\end{equation}
Expanding Eq.\ (\ref{flow+corr}) to order $|z|^{2k}$, one obtains 
an equation relating $\mean{Q}^{2k}$ to the cumulant 
$\cumul{|Q|^{2k}}$. However, 
when writing Eq.\ (\ref{denom}), we have neglected direct 
$2k$-particle correlations and autocorrelations.
As explained in Sec.\ \ref{s:b1}, both give a contribution 
of magnitude $M^{1-k}$ to the cumulant $\cumul{|Q|^{2k}}$. Thus, 
Eq.\ (\ref{flow+corr}) at order $|z|^{2k}$ is valid up to a
corrrection of order $M^{1-k}$.

\subsection{Removing autocorrelations}
\label{s:b4}

Equation (\ref{flow+corr}) can be somewhat refined. 
In the case of a $Q$ vector defined with unit weights, as in Eq.\ (\ref{qn}), 
autocorrelations can be calculated and subtracted explicitly, which is the 
purpose of this section. 

This calculation has already been done in Sec.\ \ref{s:integrated}
for the lowest orders $k=1$ and $k=2$: 
we have seen in Eq.\ (\ref{q2}) that 
diagonal terms give a contribution 1 
in the expansion of $\mean{|Q|^2}$. 
In this paper, we refer to these 
diagonal terms as ``autocorrelations''. 
Similarly, they give a contribution $-1/M$ to the fourth order 
cumulant $\cumul{|Q|^4}$, see Eq.\ (\ref{q4}) and 
Appendix~\ref{s:app-q4}.

To calculate the contribution of autocorrelations to the cumulant at an arbitrary 
order, we once again make use  of the generating function ${\cal G}_0(z)$, 
Eq.\ (\ref{genfunc1}). 
Neglecting correlations for simplicity, 
the contributions of the $M$ particles to ${\cal G}_0(z)$ 
factorize, leading to 
\begin{equation}
{\cal G}_0(z)=\mean{e^{(2x\cos\phi+2y\sin\phi)/\sqrt{M}}}_\phi^M,
\end{equation}
where we have set $z=x+iy$, and the brackets here denote an 
average over $\phi$. Assuming for simplicity that the $\phi$
distribution is isotropic, one obtains 
\begin{equation}
{\cal G}_0(z)= \left[I_0\!\left({2|z|\over\sqrt{M}}\right)\right]^M.
\end{equation} 
This is the expression of the generating function if there are only 
autocorrelations (no direct correlations, no flow). 
If there is flow, we assume that autocorrelations and 
flow give additive contributions to the cumulants, which yields 
instead of Eq.\ (\ref{flow+corr}):
\begin{equation}
\sum_{k=0}^\infty \frac{|z|^{2k}}{(k!)^2} \cumul{|Q|^{2k} } = 
\ln{\cal G}_0(z)= \ln I_0(2|z|\mean{Q })+
M\ln I_0\!\left({2|z|\over\sqrt{M}}\right).
\label{flow+corr-auto}
\end{equation}
This formula is equivalent to Eq.\ (\ref{geneq}), which we use in 
Sec.\ \ref{s:integrated2}. 
It removes exactly all autocorrelations when 
the event flow vector $Q$ is defined with unit weights, as in Eq.\ (\ref{qn}). 

\section{A generating equation for differential flow}
\label{s:app-differential}

In this appendix, we follow closely the same procedure as in Appendix
\ref{s:app-integrated}, applied to differential flow. 
In Secs.\ \ref{s:c1} and \ref{s:c2}, we first construct 
the relevant cumulants $\cumul{|Q|^{2k} Q^{*l}e^{im\psi}}$, 
as a function of the measured moments $\mean{Q^k Q^{*l}e^{im\psi}}$. 
Here, $\psi$ denotes the azimuthal angle of the particle under study
(which we call a proton), 
and $m$ the order of the harmonic measured for this particle. 
Then, we relate the cumulants to the integrated
flow $v'_m$ (Sec.\ \ref{s:c3}), and show how to 
remove autocorrelations (Sec.\ \ref{s:c4}). 

\subsection{Cluster decomposition}
\label{s:c1}

\begin{center}
\begin{figure}[ht!]
\centerline{\includegraphics*[width=0.63\linewidth]{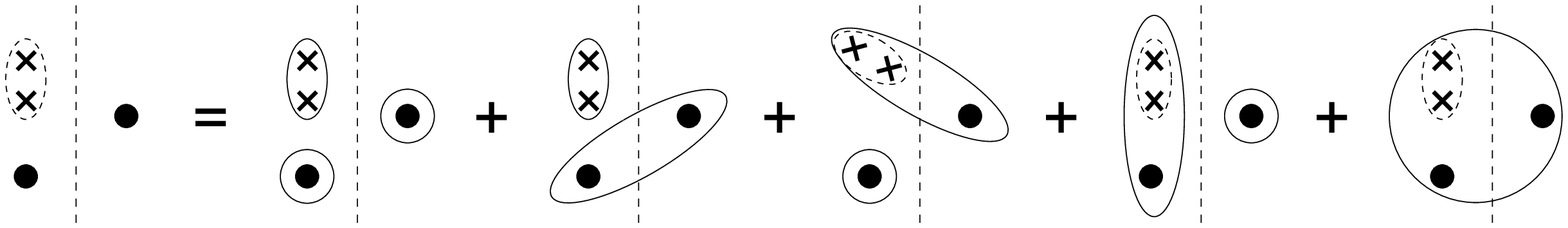}}
\medskip
\caption{Decomposition of $\mean{QQ^*e^{2i\psi}}$ in connected parts,
see Eq.\ (\ref{dcumulantsc}). 
As in Fig.\ \ref{fig:fig2p-1p}, the dot on the left (resp.\ right) 
of the dashed line stands for $Q$ (resp.\ $Q^*$). The linked 
crosses represent the proton, the number of crosses being chosen 
equal to the harmonic under study, here $m=2$.  
Circled subsets (connected diagrams) correspond to connected moments.}
\label{fig:fig2x1p-1p}
\end{figure}
\end{center}

A quantity such as $\mean{Q^k Q^{* l} e^{im\psi}}$ involves 
correlations between $k+l+1$ particles: $k+l$ ``pions'' (according to
the terminology introduced in Sec.\ \ref{s:differential}) and a proton. 
This quantity can therefore be decomposed, in the coordinate system where the 
reaction plane is fixed, into a sum of terms involving lower order 
correlations, plus a connected term of relative order $1/N^{k+l}$. 
For instance, we can write 
\begin{mathletters}
\label{dcumulants}
\begin{eqnarray}
\mean{e^{im\psi}} & = & \connex{e^{im\psi}}\label{dcumulantsa}\\
\mean{Q e^{im\psi}} & = & \connex{Q}\connex{e^{im\psi}} + \connex{Q e^{im\psi}} 
\label{dcumulantsb}\\
\mean{QQ^*e^{im\psi}} & = & \connex{Q}\connex{Q^*}\connex{e^{im\psi}} + 
\connex{QQ^*}\connex{e^{im\psi}} + \connex{Q}\connex{Q^* e^{im\psi}} + 
\connex{Q^*}\connex{Q e^{im\psi}} + \connex{QQ^*e^{im\psi}}, \label{dcumulantsc}
\end{eqnarray}
\end{mathletters}
where, in the third equation, the last term is of order $1/N^2$ relative to the first one. 
Such decompositions can be represented diagrammatically, in a way 
similar to the decomposition of $\mean{Q^k Q^{* l}}$ 
in Appendix \ref{s:app-integrated}. 
We choose to represent the proton by $m$ crosses on the left, 
for reasons which will become clear below, when we 
consider the specific case of an isotropic source. 
For instance, Eq.\ (\ref{dcumulantsc}) can be
represented diagrammatically by Fig.\ \ref{fig:fig2x1p-1p}. 

In order to express in a compact way the relations between 
the moments $\mean{Q^k Q^{* l} e^{im\psi}}$ and the corresponding
connected moments $\connex{Q^k Q^{* l} e^{im\psi}}$, we introduce 
the following generating function 
\begin{equation}
\label{defgm}
{\cal G}_m(z)=\mean{e^{z^*Q+zQ^*}e^{im\psi}}
=\sum_{k,l} {z^{*k} z^{ l}\over k!\, l!} \mean{Q^k Q^{* l}e^{im\psi}},  
\end{equation}
Expanding ${\cal G}_m(z)$ to order $z^{*k} z^{ l}$, one obtains 
all the moments $\mean{Q^k  Q^{* l} e^{im\psi}}$. 
In order to obtain the generating function of the connected moments, 
we note that each diagram in Fig.\ \ref{fig:fig2x1p-1p} 
can be written as the product of 
a connected diagram containing  the crosses, i.e.\ the proton, 
times an arbitrary diagram (not necessarily connected) involving 
only pions, which corresponds to the terms 
$\mean{Q^k Q^{* l}}$ considered in Appendix 
\ref{s:app-integrated}. 
For instance, using Eqs.\ (\ref{qcumulantsa}) and (\ref{qcumulantsc}),
one can rewrite Eq.\ (\ref{dcumulantsc}) as 
\begin{equation}
\label{semi}
\mean{QQ^*e^{im\psi}} = \mean{QQ^*}\connex{e^{im\psi}} + 
\mean{Q}\connex{Q^* e^{im\psi}} + \mean{Q^*}\connex{Q e^{im\psi}} + 
\connex{QQ^*e^{im\psi}}. 
\end{equation}
Therefore, the generating function of the diagrams with pions and protons 
$\mean{e^{z^*Q + z Q^* + im\psi} }$ 
is the product of the generating 
function of graphs with only pions, i.e.\ ${\cal G}_0(z)$ 
defined in Eq.\ (\ref{genfunc1}), by the generating function of 
connected graphs with pions and protons. 
This latter is therefore: 
\begin{equation}
{\cal C}_m(z)=\sum_{k,l} {z^{*k}z^l\over k!\, l!}
\connex{Q^k Q^{* l} e^{im\psi}}
 \equiv{ {\cal G}_m(z)\over {\cal G}_0(z)}=
\frac{\mean{e^{z^*Q + z Q^* + im\psi} }}{\mean{e^{z^*Q + z Q^*} 
}}
\label{genfuncdiff1}
\end{equation}
As in Eq.\ (\ref{logG}), the normalization coefficient $1/k!\,l!$ has been 
chosen so that $\connex{Q^k Q^{* l} e^{im\psi}}$ appears with a unit 
coefficient in the expansion of $\mean{Q^k Q^{* l} e^{im\psi}}$.

\subsection{Isotropic source}
\label{s:c2}

We now consider the particular case of an isotropic source,
without flow. 
Then the moment $\mean{Q^k Q^{* l} e^{im\psi} }$ 
vanishes when $k+m \neq l$, and so do the corresponding connected 
parts. This is the reason why we chose to represent the proton with $m$ 
crosses: in the isotropic case, only diagrams with the same number of 
points (crosses and dots) on each side of the dashed line do not vanish. 

Expanding the generating function (\ref{defgm}) and keeping
only the nonvanishing terms, one finds 
\begin{eqnarray}
{\cal G}_m(z)&=&\sum_{k=0}^\infty \frac{|z|^{2k} z^m}{k!\,(k+m)!} 
\mean{|Q|^{2k} {Q^*}^m e^{im\psi} }\cr
&=&\mean{I_m(2|zQ|) \left(\frac{zQ^*}{|zQ|}\right)^m e^{im\psi}},
\end{eqnarray}
where $I_m$ is the modified Bessel function of order $m$. 

We define the cumulants $\cumul{|Q|^{2k} {Q^*}^m e^{im\psi}}$
so that they coincide with the connected moments in 
Eq.\ (\ref{genfuncdiff1}) when the source is 
isotropic. Using Eq.\ (\ref{genfunc2}), this gives
\begin{eqnarray}
{\cal C}_m(z) &=& 
\sum_{k=0}^\infty \frac{|z|^{2k} z^m}{k!\,(k+m)!} 
\cumul{|Q|^{2k} {Q^*}^m e^{im\psi}} \cr
&\equiv& \frac{\mean{I_m(2|zQ|) 
\left(\frac{z Q^*}{|zQ|}\right)^m e^{im\psi}
}}{\mean{I_0(2|zQ|) }}.
\label{genfuncdiff2}
\end{eqnarray}
This equation is equivalent to Eq.\ (\ref{gendiffcumul}), setting 
$x=|z|$. 

\subsection{Flow}
\label{s:c3}

Finally, we turn to the more general case of collisions with flow. 
Neglecting for simplicity nonflow correlations between particles, 
the generating function (\ref{defgm}) becomes
\begin{equation}
{\cal G}_m(z) = e^{(z+z^*) \mean{Q }}v'_m.
\label{denomm}
\end{equation}
As explained in Sec.\ \ref{s:b3}, this 
quantity is measured in the laboratory coordinate system, 
therefore one must replace  $z$ by $ze^{i\phi_R}$ and 
average the new expression over all possible $\phi_R$. 
That yields
\begin{eqnarray}
{\cal G}_m(z) &=& 
v'_m\displaystyle \int_0^{2\pi} \! e^{(ze^{i\phi_R} + z^*e^{-i\phi_R}) \mean{
Q }} e^{im\phi_R} \, {{\rm d}\phi_R\over 2\pi} \cr 
&=& I_m(2|z|\mean{Q }) \, \left( 
\frac{z}{|z|}\right)^m v'_m. 
\end{eqnarray}
Using Eq.\ (\ref{genfuncflow}), the generating function of cumulants 
(\ref{genfuncdiff1}) takes the form 
\begin{equation}
{\cal C}_m(z) = \frac{I_m(2|z|\mean{Q })} {I_0(2|z|\mean{Q })} 
\, \left(\frac{z}{|z|}\right)^m v'_m.
\label{genfuncdiffflow}
\end{equation}
Gathering Eqs.\ (\ref{genfuncdiff2}) and 
(\ref{genfuncdiffflow}), we obtain 
\begin{equation}
\sum_{k=0}^\infty \frac{|z|^{2k} z^m}{k!\,(k+m)!} 
\cumul{|Q|^{2k} {Q^*}^m e^{im\psi}} 
={\cal C}_m(z) =
\frac{I_m(2|z|\mean{Q })} {I_0(2|z|\mean{Q })} 
\, \left(\frac{z}{|z|}\right)^m v'_m.
\label{flow+corr-diff}
\end{equation}
Expanding this equation to order $|z|^{2k}z^m$, one obtains 
a proportionality relation between the cumulant 
$\cumul{|Q|^{2k} {Q^*}^m e^{im\psi}}$ and 
$\mean{Q}^{2k+m} v'_m$. Having measured independently 
the integrated flow $\mean{Q}$, one thus obtains the differential 
flow $v'_m$ from the cumulant. As discussed in 
Sec.\ \ref{s:differential}, the corresponding error from 
nonflow correlations is of order $(\mean{Q}\sqrt{M})^{-k-(m/2)}$. 

\subsection{Removing autocorrelations}
\label{s:c4}

In the case when the ``proton'' is included in the construction 
of the event flow vector $Q_n$, i.e.\ if $\psi$ is one of the angles 
$\phi_j$ in Eq.\ (\ref{qn}), the resulting autocorrelations can 
be removed at the level of the generating function ${\cal C}_m(z)$ 
in Eq.\ (\ref{genfuncdiff2}): 
this subtraction is similar to that performed in Sec.\ \ref{s:b4}
for the integrated flow. 

Neglecting correlations for simplicity, the generating function 
of the cumulants, defined by Eq.\ (\ref{genfuncdiff1}), becomes 
\begin{equation}
{\cal C}_m(z)={\mean{e^{(2x\cos\psi+2y\sin\psi+im\psi)/\sqrt{M}}}_\psi \over
\mean{e^{(2x\cos\psi+2y\sin\psi)/\sqrt{M}}}_\psi},
\end{equation}
where we have set $z=x+iy$, and the brackets denote an 
average over $\psi$. Assuming for simplicity that the $\psi$
distribution is isotropic, one obtains 
\begin{equation}
{\cal C}_m(z)= {I_m\left({2|z|/\sqrt{M}}\right)\over
I_0\left({2|z|/\sqrt{M}}\right)}\left({z\over |z|}\right)^m .
\end{equation} 
This is the value of the generating function if there are only 
autocorrelations.
If there is flow in addition, we assume that the contributions 
of autocorrelations and flow are additive. 
Equation (\ref{flow+corr-diff}) is then replaced by 
\begin{equation}
\sum_{k=0}^\infty \frac{|z|^{2k} z^m}{k!\,(k+m)!} 
\cumul{|Q|^{2k} {Q^*}^m e^{im\psi}} =
\left(\frac{I_m(2|z|\mean{Q })} {I_0(2|z|\mean{Q })} 
 v'_m +{I_m\left({2|z|/\sqrt{M}}\right)\over
I_0\left({2|z|/\sqrt{M}}\right)}\right)
\, \left(\frac{z}{|z|}\right)^m. 
\label{flow+corr-diff-auto}
\end{equation}
This equation is equivalent to Eq.\ (\ref{geneqdiff}), 
setting $x=|z|$. 
This formula removes exactly all autocorrelations when 
the vector $Q_n$ is defined with unit weights.

\section{Interpolation formulas}
\label{s:interpol}

In this Appendix, we give interpolation methods
to calculate numerically the cumulants from 
their generating functions.

\subsection{Integrated flow}
\label{s:interpoli}

The cumulants used for the measurement of the integrated 
flow are defined by Eq.\ (\ref{cumuli}). 
In order to compute numerically the cumulants 
$\langle\!\langle |Q^{2k'}|\rangle\!\rangle$ 
for $k'=1\dots k$ from the generating function, 
one can for instance tabulate the generating function 
at the following points:
\begin{equation}
G_{p,q}\equiv \log{\cal G}_0
\left(r_0\sqrt{p}\,\cos {2\,q\,\pi\over q_{\rm max}},
r_0\sqrt{p}\,\sin{2\,q\,\pi\over q_{\rm max}}\right)
\end{equation}
for $p=1,\dots,k$ and $q=0,\dots,q_{\rm max}-1$. In this equation, 
$r_0$ is a real number which should be chosen small enough 
for the series expansion to converge rapidly, 
typically $r_0\sim 0.1$, and $q_{\rm max}$ is the number of angles 
at which the generating function is evaluated, which should 
satisfy the condition $q_{\rm max}>2k$. 

One then averages over the angle, thereby eliminating nonisotropic
terms up to order $|z|^{2k}$:
\begin{equation}
G_{p}\equiv {1\over q_{\rm max}}\sum_{q=0}^{q_{\rm max}-1}G_{p,q}.
\end{equation}
Then, the $G_p$, with $p=1,\dots,k$, 
are related to the cumulants 
$\langle\!\langle |\bar Q|^{2k'}\rangle\!\rangle$ with $k'=1,\dots,k$ 
by the following linear system of equations:
\begin{equation}
G_p=\sum_{k'=1}^{k} \cumul{|\bar Q|^{2k'}}
{r_0^{2k'}\over (k'!)^2}\, p^{k'} \quad 1 \leq p \leq k.
\end{equation}
For practical purposes, it is enough to take $k=3$, as explained 
in Sec.\ \ref{s:statistical}. In this case, 
the solution of the above system reads 
\begin{eqnarray}
\label{into6}
\cumul{|\bar Q|^2}&=&{1\over r_0^2}\,
\left(3\, G_1-{3\over 2}G_2+{1\over 3}G_3\right)
,\cr
\cumul{|\bar Q|^4}&=&{2\over r_0^4}\,
\left(-5\, G_1+4\,G_2-\,G_3\right)
,\cr
\cumul{|\bar Q|^6}&=&{6\over r_0^6}\,
\left(3\, G_1-3\,G_2+G_3\right).
\end{eqnarray}

\subsection{Differential flow}
\label{s:interpold}

The cumulants used for the measurement of the harmonic $v'_m$ 
are defined from the generating function by Eq.\ (\ref{devcm}). 
In order to compute numerically the cumulants 
$\langle\!\langle |Q|^{2k'}Q^{*m}e^{im\psi}\rangle\!\rangle$ 
for $k'=0,\dots,k$, 
from the generating function, we first 
tabulate the real and imaginary parts 
of the generating function, defined by Eq.\ (\ref{cxy}),  
at the following points:
\begin{eqnarray}
X_{p,q}\equiv {\Re}\left[{\cal C}_{m}\! \left(r_0\sqrt{p}\,\cos
\frac{2\,q\,\pi}{q_{\rm max}},
r_0\sqrt{p}\,\sin\frac{2\,q\,\pi}{q_{\rm max}}\right)\right]\cr
Y_{p.q}\equiv {\Im}\left[{\cal C}_{m}\! \left(r_0\sqrt{p}\,
\cos \frac{2\,q\,\pi}{q_{\rm max}},
r_0\sqrt{p}\,\sin\frac{2\,q\,\pi}{q_{\rm max}}\right)\right]
\end{eqnarray}
for $p=1,\dots,k+1$ and $q=0,\dots, q_{\rm max}-1$. The number of angles 
$q_{\rm max}$ must satisfy the condition $q_{\rm max}>2(k+m)$, 
as we see below.

One then multiplies ${\cal C}_m(z)$ by $z^{*m}$, takes the real part and 
averages over azimuthal angles. Provided $q_{\rm max}$ is large
enough, one thus eliminates all nonisotropic terms up to  
order $z^{*k}z^{k+m}$ in the generating function:
\begin{equation}
C_{p}\equiv {\left (r_0\sqrt{p}\right)^m 
\over q_{\rm max}}\sum_{q=0}^{q_{\rm max}-1}
\left[\cos\left(m\frac{2\,q\,\pi}{q_{\rm max}}\right) \, X_{p,q}+
\sin\left(m\frac{2\,q\,\pi}{q_{\rm max}}\right) \,Y_{p,q}\right]. 
\end{equation}
Then, the values of $C_{p}$ for $p=1,\dots k+1$ are related 
to the cumulants  $\langle\!\langle|Q|^{2k'}Q^{*m}e^{im\psi}\rangle\!\rangle$ for 
$k'=0,\dots,k$ by the following linear system of equations:
\begin{equation}
C_{p}=\sum_{k'=0}^{k}\cumul{|\bar Q|^{2k'}\bar Q^{*m}e^{im\bar\psi}}
{r_0^{2(k'+m)} p^{k'+m}\over k'!(k'+m)!}, 
\quad 1 \leq p \leq k+1.
\end{equation}
Taking $k=1$ is sufficient for most purposes, as shown in 
Sec.\ \ref{s:differential3}. 
For $m=1$,  the solution of this system is 
\begin{eqnarray}
\cumul{\bar Q^{*}e^{i\bar\psi}}&=&
{1\over r_0^2}\left(2\, C_1-{1\over 2}\, C_2\right),\cr
\cumul{|\bar Q|^{2}\bar Q^{*}e^{i\bar \psi}}
&=&{1\over r_0^4}\left( -2\, C_1+ C_2\right),
\end{eqnarray}
while for $m=2$, 
\begin{eqnarray}
\cumul{\bar Q^{*2}e^{2i\bar\psi}}
&=&{1\over r_0^4}\left(4\, C_1-{1\over 2}\, C_2\right),\cr
\cumul{|\bar Q|^{2}\bar Q^{*2}e^{2i\bar\psi}}
&=&{1\over r_0^6}\left( -6\,C_1+{3\over 2}\, C_2\right).
\end{eqnarray}


\begin{thebibliography}{99}

\bibitem{ollitrault98} 
For a review, see J.-Y. Ollitrault, 
Nucl. Phys. {\bf A638}, 195c (1998). 

\bibitem{danielewicz85}
P. Danielewicz and G. Odyniec, 
Phys. Lett. {\bf 157B}, 146 (1985).

\bibitem{dinh} 
P. M. Dinh, N. Borghini, and J.-Y. Ollitrault, 
Phys. Lett. {\bf B477}, 51 (2000).

\bibitem{borghini}
N. Borghini, P. M. Dinh, and J.-Y. Ollitrault, 
Phys. Rev. {\bf C62}, 034902 (2000).

\bibitem{danielewicz88}
P. Danielewicz {\it et al.\/}, 
Phys. Rev. {\bf C38}, 120 (1988).

\bibitem{Sorge} H. Sorge, Phys. Rev. Lett. {\bf 82} (1999) 2048. 

\bibitem{Volo99} S. A. Voloshin and A. M. Poskanzer, 
Phys. Lett. {\bf B474} (2000) 27. 

\bibitem{Bravina} L. V. Bravina, A. Faessler, C. Fuchs,
and E. E. Zabrodin, Phys. Rev. {\bf C61} (2000) 064902. 

\bibitem{poskanzer98}
A. M. Poskanzer and S. A. Voloshin, 
Phys. Rev. {\bf C58}, 1671 (1998).

\bibitem{WA93} M. M. Aggarwal {\it et al.\/}, WA93 Collaboration,
Phys. Lett. {\bf B403} (1997) 390; W.H. van Heeringen, Ph.D. Thesis, 
ISBN 90-393-1325-3, Universiteit Utrecht (1996). 

\bibitem{STAR00}
K. H. Ackermann {\it et al.\/} (STAR Collaboration), 
preprint nucl-ex/0009011.

\bibitem{Jiang92} 
J. Jiang {\it et al.\/}, 
Phys. Rev. Lett. {\bf 68}, 2739 (1992). 

\bibitem{voloshin96}
S. A. Voloshin and Y. Zhang, 
Z. Phys. {\bf C70}, 665 (1996).

\bibitem{barrette94}
J. Barrette {\it et al.\/} (E877 Collaboration), 
Phys. Rev. Lett. {\bf 73}, 2532 (1994). 

\bibitem{lenkeit} 
B. Lenkeit for the CERES Collaboration, 
Nucl. Phys. {\bf A661}, 23c (1999).

\bibitem{NA49unpub} 
A. M. Poskanzer and S. A. Voloshin, unpublished. 

\bibitem{na49}
H.~Appelsh\"auser {\it et al.\/} (NA49 Collaboration), 
Phys. Rev. Lett. {\bf 80}, 4136 (1998). 

\bibitem{E877} 
J. Barrette {\it et al.\/} (E877 Collaboration), 
Phys. Rev. {\bf C55}, 1420 (1997). 

\bibitem{hill} 
T. L. Hill, {\it Statistical Mechanics}, 
(McGraw-Hill, NY, 1956), chapter 5. 

\bibitem{vanKampen}
N.~G.~van Kampen, {\it Stochastic processes in physics and chemistry} 
(North-Holland, Amsterdam, 1981).

\bibitem{carruthers89} 
P. Carruthers and I. Sarcevic, 
Phys. Rev. Lett. {\bf 63}, 1562 (1989). 

\bibitem{giovannini77} 
A. Giovannini and G. Veneziano, 
Nucl. Phys. {\bf B130}, 61 (1977). 

\bibitem{eggers93} 
H. C. Eggers, P. Lipa, P. Carruthers, and B. Buschbeck, 
Phys. Lett. {\bf B301}, 298 (1993). 

\bibitem{ollitrault92} 
J.-Y. Ollitrault, Phys. Rev. {\bf D46}, 229 (1992); 
Phys. Rev. {\bf D48}, 1132 (1993). 

\bibitem{wilson92} 
W. K. Wilson, R. Lacey, C. A. Ogilvie, and G. D. Westfall, 
Phys. Rev. {\bf C45}, 738 (1992). 

\bibitem{ollitrault95}
J.-Y.~Ollitrault, 
Nucl.~Phys.~{\bf A590}, 561c (1995).

\bibitem{na50}
P. Saturnini (NA50 Collaboration), 
Nucl. Phys. {\bf A661}, 345c (1999). 

\bibitem{poskanzer99} 
A. M. Poskanzer and S. A. Voloshin (NA49 Collaboration) 
Nucl. Phys. {\bf A661}, 341c (1999). 

\bibitem{danielewicz83}
P. Danielewicz and M. Gyulassy, 
Phys. Lett. {\bf 129B}, 283 (1983). 

\bibitem{gustafsson84}
H. Gustafsson {\it et al.\/}, 
Phys. Rev. Lett. {\bf 52}, 1590 (1984). 

\bibitem{doss87} 
K. G. R. Doss {\it et al.\/}, 
Phys. Rev. Lett. {\bf 59}, 2720 (1987).

\bibitem{ollitrault97}
J.-Y. Ollitrault, ``A method of reconstructing azimuthal distributions 
in heavy ion collisions,''
in {\it XXV International Symposium on Multiparticle Dynamics\/}, 
Stara Lesna, Slovakia, 11-16 september 1995, 
Bruncko D., Sandor L., Urban J. eds.,  
(World Scientific, 1996), pp. 290-296; 
J.-Y. Ollitrault, Los Alamos preprint nucl-ex/9711003, unpublished. 

\bibitem{E877bis} 
J. Barrette {\it et al.\/} (E877 Collaboration), 
Phys. Rev. {\bf C56}, 3254 (1997);
Phys. Rev. {\bf C59}, 884 (1999).

\bibitem{inprogress}
N. Borghini, P. M. Dinh, and J.-Y. Ollitrault, 
in preparation. 

\bibitem{domb-green}
C.~Domb and M.~S.~Green Eds., {\it Phase transitions and critical phenomena, 
vol.3: Series expansions for lattice models} (Academic Press, New York, 1974).

\end{thebibliography}
\end{document}